\documentclass[sort&compress]{elsarticle}
\usepackage{amsmath,amssymb,amsthm,mathtools,bm}
\DeclareMathOperator*{\erf}{erf}
\usepackage{booktabs}
\usepackage{caption,subcaption}
\usepackage{xcolor,graphicx}
\usepackage{hyperref}
\usepackage[capitalize]{cleveref}
\usepackage{enumitem}

\usepackage{tikz,pgfplots}
\usetikzlibrary{shapes.geometric, arrows, patterns,fit,decorations.pathreplacing}
\usepgfplotslibrary{external}
\definecolor{carnelian}{RGB}{179,27,27}

\newlength\figurewidth

\newcommand{\includetikz}[2]{%
  \includegraphics{tikz/figs/#2}%
}%
\DeclareRobustCommand{\reftikz}[2]{%
  \includegraphics{tikz/figs/#1}%
}%
\title{A volume of fluid framework for interface-resolved simulations of vaporizing liquid-gas flows}
\author[VT]{John Palmore Jr.\corref{cor}}
\ead{palmore@vt.edu}
\cortext[cor]{Corresponding author.}
\address[VT]{Department of Mechanical Engineering, Virginia Tech, Blacksburg, VA, USA.}
\author[cornell]{Olivier Desjardins}
\ead{olivier.desjardins@cornell.edu}
\address[cornell]{Sibley School of Mechanical \& Aerospace Engineering, Cornell University, Ithaca, NY, USA.}

\begin{document}

\begin{abstract}
  This work demonstrates a computational framework for simulating vaporizing, liquid-gas flows. It is developed for the general vaporization problem \cite{Villegas2016} which solves the vaporization rate based as from the local thermodynamic equilibrium of the liquid-gas system. This includes the commonly studied vaporization regimes of film boiling and isothermal evaporation. The framework is built upon a Cartesian grid solver for low-Mach, turbulent flows \cite{Desjardins2008b} which has been modified to handle multiphase flows with large density ratios \cite{Desjardins2013}. Interface transport is performed using an unsplit volume of fluid solver \cite{Owkes2014}. A novel, divergence-free extrapolation technique is used to create a velocity field that is suitable for interface transport. Sharp treatments are used for the vapor mass fractions and temperature fields \cite{Ma2013}. The pressure Poisson equation is treated using the Ghost Fluid Method \cite{Liu2000}. Interface equilibrium at the interface is computed using the Clausius-Clapeyron relation, and is coupled to the flow solver using a monotone, unconditionally stable scheme.
  
  It will be shown that correct prediction of the interface properties is fundamental to accurate simulations of the vaporization process. The convergence and accuracy of the proposed numerical framework is verified against solutions in one, two, and three dimensions. The simulations recover first order convergence under temporal and spatial refinement for the general vaporization problem. The work is concluded with a demonstration of unsteady vaporization of a droplet at intermediate Reynolds number.
  %% vaporization in complex flow. First the unsteady vaporization of a droplet at intermediate Reynolds number. Second, the vaporization of an atomizing liquid jet.
\end{abstract}


\maketitle
\section{Introduction}
Several common engineering systems involve the flow of liquids and gases. For many of these flows, it is often the case that the primary objective of these systems is to exchange heat or mass between the two phases. Some common systems include heat exchangers, spray coolers, bubble column reactors, and spray combustors. In these systems heat and mass transfer occurs primarily through the process of phase change. As such, understanding phase change is an important step in better understanding the dynamics of these engineering systems. This paper focuses on liquid-to-gas phase change, here called vaporization. For many of these problems, it is appropriate to conceptualize the flow as consisting of a continuous carrier phase with dispersed droplets. Several reviews have been written on the theory of the vaporization of droplets \cite{Law1982,Sirignano1983,Dhir1998}. These studies were focused primarily on either vaporization for idealized flows. The study of the interaction of vaporization and turbulent flows remains an open problem.
 
Recently, experiments have been performed studying vaporization and combustion of droplets in turbulent flows (e.g., the review by Birouk and Gokalp \cite{Birouk2006}), however a fundamental understanding of the mechanism of vaporization-turbulence interaction is lacking. In particular, the question of which scales of turbulence are most important to the interaction is yet to be answered. Direct numerical simulation (DNS) can provide an alternative avenue to study this problem.

In this context, DNS are taken to be those simulations that solve the dynamics of the flow from first principles, i.e., directly from the laws for conservation of mass, momentum, and energy. Vaporization is represented as a set of coupled matching conditions at the liquid-gas interface which ensure the conservative transfer of mass, momentum, and energy across the interface. According to Sirignano \cite{Sirignano2010}, a multiphase flow simulation can be considered a true DNS when it resolves both the gas film around the droplet and the internal droplet flow. Unfortunately, due to the sharp discontinuities in pressure, velocity, and density at the interface, the stable numerical simulation of this problem is difficult.

Only recently have numerical methods capable of performing resolved simulations of vaporization have been discussed \cite{Son1998,Welch2000,Tanguy2007,Gibou2007,Schlottke2008,Hardt2008,Ma2013,Villegas2016,Anumolu2018a}. The aforementioned numerical frameworks have been used to study film boiling, vaporizing bubbles rising in liquid, and the classical \(d^2\) law. However, the studies are performed under several assumptions regarding the regime of vaporization and the flow conditions.

First, simulations are typically performed within one of two limiting cases of vaporization: an evaporation limit wherein the vaporization rate is assumed to be limited by the local values in the vapor species \cite{Tanguy2007,Schlottke2008}; and a boiling limit wherein the vaporization rate is assumed to be limited by the local values in temperature \cite{Son1998,Welch2000,Gibou2007,Hardt2008,Anumolu2018a}. Few authors have studied vaporization without the imposition of these constraints \cite{Schlottke2008,Ma2013,Villegas2016}. All of the frameworks mentioned so far used incompressible solvers, and assumed constant thermophysical properties. Furthermore, most of the studies were performed in two dimensions which inherently neglects many important physical processes in turbulent flows. %% such as vortex stretching, the turbulent kinetic energy cascade, and turbulent mixing.

The literature also contains very little discussion of the convergence and stability of the numerical methods used in the simulations. The current work demonstrates a numerical framework for vaporization in three dimensions, and develops a suite of tests to study the stability and convergence of the flow solver. In developing the framework, it will be shown that traditional semi-implicit treatments of scalar transport can lead to stability issues for large timesteps. %% It will also be seen that an accurate implementation of the interface equilibrium conditions is required in order to recover a convergent scheme.

The work is organized as follows. \Cref{sec:physics} gives the physical description of the problem and outlines the mathematical model of the flow. \Cref{sec:numerics} details the algorithms used in the work. \Cref{sec:results} verifies the numerical algorithm against known analytical solutions in one dimension. \Cref{sec:2D} performs a verification study in two dimensions, and it demonstrates the temporal accuracy and stability of the solver. \Cref{sec:3D} verifies the solver against the well know \(d^2\) solution in three dimensions. Finally, demonstrations of unsteady droplet vaporization (\cref{sec:cross}) are shown. %% and vaporization of an atomizing jet (\cref{sec:jet}) are demonstrated.
\section{Physical Description}\label{sec:physics}
This section details the mathematical equations used to describe the vaporization process. The process is described by the conservation equations for low-Mach flow, i.e., conservation of mass, momentum, energy, and chemical species in each phase. Thermodynamic equilibrium at the interface is resolved using the Clausius-Clapeyron relation. As an initial assumption, a single component fuel is used, and the fuel vapor does not react with the surrounding gas. Both the liquid and the gas are assumed to be of constant density. The exposition of the governing equations follows.

The motion of a Newtonian fluid is governed by the Navier-Stokes equations,
\begin{subequations}
\begin{align}
  \frac{\partial\left(\rho\bm u\right)}{\partial t}+\nabla\cdot\left(\rho\bm u\otimes\bm u\right)=-\nabla p+\nabla\cdot\left(\mu\bm {\mathcal S}\right), \text{ where} \label{eq:NS1}  \\
  \bm {\mathcal S}=\left(\nabla\bm u+{\nabla\bm u}^\top-\frac{2}{3}\left(\nabla\cdot\bm u\right)\bm{\mathcal I}\right), \label{eq:NS2}
\end{align}\label{eq:NS}
\end{subequations}
which is statement of the conservation of momentum. Here, \(\rho\) and \(\mu\) are the fluid density and dynamic viscosity; \(\bm u\) is the velocity; and \(p\) is the pressure. The identity tensor is represented with \(\bm{\mathcal I}\). \Cref{eq:NS} must be coupled with the continuity equation, here expressed as
\begin{align}
  \frac{\partial\rho}{\partial t}+\nabla\cdot\left(\rho\bm u\right)=0,\label{eq:cont}
\end{align}
which is an expression of the conservation of mass.

In the gas phase, conservation of mass must be supplemented with an equation for conservation of chemical species, 
\begin{align}
  \frac{\partial\left(\rho Y\right)}{\partial t}+\nabla\cdot\left(\rho Y\bm u\right)=\nabla\cdot\left(\rho D\nabla Y\right),\label{eq:vapor_frac}
\end{align}
where the vapor mass fraction and its diffusivity are represented using \(Y\) and \(D\), respectively. As the problem is nonreactive, remaining inert gas can be expressed as \(1-Y\).

The equation for conservation of energy can be expressed as
\begin{align}
  \frac{\partial\left(\rho C_p T\right)}{\partial t}+\nabla\cdot\left(\rho C_p T\bm u\right)=\nabla\cdot\left(k\nabla T\right)+\frac{Dp}{Dt}+\frac{\mu}{2}\bm {\mathcal S}:\bm {\mathcal S},\label{eq:energy_fake}
\end{align}
where \(T\) is the fluid temperature, \(C_p\) is the specific heat under constant pressure, and Fourier's law is used for conduction (\(k\) is the thermal conductivity). The notation, \(D/Dt\) represents the  Lagrangian derivative, i.e.,
\begin{align}
  \frac{Dp}{Dt}=\frac{\partial p}{\partial t}+\bm u\cdot\nabla p,
\end{align}
and double tensor contraction expands as
\begin{align}
  \bm {\mathcal S}:\bm {\mathcal S}=S_{ij}S_{ij},
\end{align}
using Einstein summation convention.

However, the full energy equation is not used in this work. Consistent with the low-Mach assumption, we neglect those terms of the energy equation that are quadratic in the Mach number (or equivalently the velocity), yielding
\begin{align}
  \frac{\partial\left(\rho C_p T\right)}{\partial t}+\nabla\cdot\left(\rho C_p T\bm u\right)=\nabla\cdot\left(k\nabla T\right).\label{eq:energy}
\end{align}
In deriving \cref{eq:energy} it was assumed that the pressure scales as the dynamic pressure, \(\frac{1}{2}\rho|\bm u|^2\). The resulting equation neglects all pressure and viscous work on the system and, accordingly, acoustic effects.
%% This includes neglecting acoustic effects. (See the derivation by \cite{Majda1985}).
%% which can be expressed as \(\frac{Dp}{Dt}\) and \(\nabla\bm u:\mu\bm {\mathcal S}\), respectively
\subsection{Phase Change Strategy}
\begin{figure}
  \centering
  \begin{subfigure}{0.49\linewidth}
    \centering
    \includetikz{\linewidth}{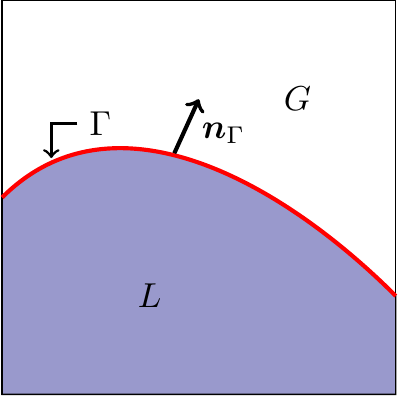}
    \caption{Geometry of the problem.}\label{fig:interface_geom}
  \end{subfigure}
  \begin{subfigure}{0.49\linewidth}
    \centering
    \includetikz{\linewidth}{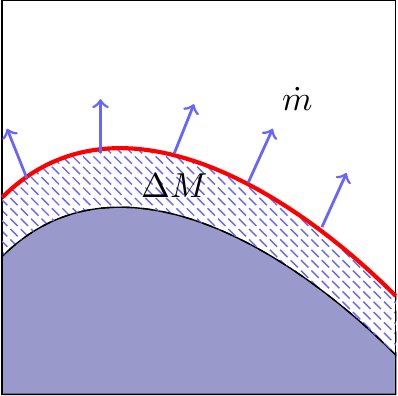}
    \caption{Illustration of the vaporization process}\label{fig:interface_mdot}
  \end{subfigure}
  \caption{Geometry near the liquid gas interface. \Cref{fig:interface_geom}: The liquid and gas regions are labeled as \(L\) and \(G\), respectively, whereas the interface is \(\Gamma\). The local interface normal vector, \(\bm n_\Gamma\) is shown as well. \Cref{fig:interface_mdot}: Over a time, \(\Delta t\), vaporization at the rate, \(\dot m\), causes a change in liquid mass of \(\Delta M\)}\label{fig:interface}%=\int\limits_t^{t+\Delta t}\int\limits_{\Gamma\cap{\Omega_i}}\dot mdSdt\).}\label{fig:interface}
\end{figure}

The conservation equations \cref{eq:NS,,eq:cont,eq:vapor_frac,eq:energy} do not offer a complete description of the flow for multiphase problems. A sample domain is shown in \cref{fig:interface_geom}. To completely describe the system, the interface location and interface matching conditions must be known. Label the liquid and gas portions of the domain using \(L\) and \(G\), respectively. The liquid-gas interface is then defined as the intersection of these two regions, and is referred to with \(\Gamma\). On the interface, an outward facing normal, \(\bm n_\Gamma\), is defined pointing away from the liquid. An vaporization mass flux, \(\dot m\), forms normal to the interface, and removes an amount of mass \(\Delta M\) from the liquid (\cref{fig:interface_mdot}).

The notation \(\cap\) will be used to refer to the intersection of two sets, so that \(\Gamma=L\cap G\). The notation \(\cup\) will be used for the union of two sets, so that the entire domain is given by \(L\cup G\). In \cref{fig:interface}, the liquid region is represented using a dark color, the gas is represented using white, and the interface is represented as a thick line. This convention is used throughout the remainder of this work.

Because only two phases exist in this problem, \(\Gamma\) can be identified using the liquid indicator function,
\begin{align}
  f(\bm x,t)=\left\{
  \begin{array}{cc}
    1 & \text{if }\bm x \in L,\\
    0 & \text{otherwise.}
  \end{array}
  \right.
\end{align}
The interface lies where \(f\) passes between \(0\) and \(1\), or equivalently when \(\nabla f\) is nonzero. In practice, numerical schemes transport the liquid volume fraction, \(\alpha\), which is the volume average of \(f\) in each computational cell.

\subsection{Interfacial Conditions}
Consider the local velocity of interface motion, \(\bm {u_S}\). It follows from the principle of conservation of mass that,
\begin{align}
  \bm n_\Gamma\cdot\rho_G\left(\bm u_G-\bm u_S\right)=\bm n_\Gamma\cdot\rho_L\left(\bm u_L-\bm u_S\right)=\dot m.\label{eq:interface_cont}
\end{align}
This expresses the fact that the flux of mass into the gas from the interface is the same as the flux from the liquid into the interface. This flux is denoted \(\dot m\). It satisfies
\begin{align}
  \frac{d}{dt}\int\limits_L\rho_LdV=-\int\limits_{\Gamma}\dot mdS,\label{eq:mass_cont}
\end{align}
hence, \(\dot m\) is the local vaporization rate per unit area. (By convention, integrals of type \(dS\) refer to surface integrals and of type \(dV\) refer to volume integrals.) This also implies a transport equation for the liquid volume fraction in the form,
\begin{align}
  \frac{\partial\rho_L\alpha}{\partial t}+\nabla\cdot\left(\rho_L\alpha\bm u\right)=-\dot m\delta_\Gamma,\label{eq:VOF}
\end{align}
which is derived from substituting the definitions of \(f\) and \(\alpha\) into \cref{eq:mass_cont} and using the Reynolds transport theorem. Here \(\delta_\Gamma\) is the interfacial surface area density which can be defined for a region \(\Omega\) as,
\begin{align}
  \delta_\Gamma =\int\limits_{\Gamma\cap{\Omega}}dS\left/\int\limits_{\Omega}dV\right..
\end{align}

\Cref{eq:interface_cont} can also be expressed as an interfacial compatibility condition between the velocities of the two phases. Defining a jump notation as
\begin{align}
  \left[\phi\right]_\Gamma=\phi_G-\phi_L,
\end{align}
the interface condition, \cref{eq:interface_cont} can be written as
\begin{align}
  \left[\bm n_\Gamma\cdot\bm u\right]_\Gamma=\dot{m}\left[\frac{1}{\rho}\right]_\Gamma.\label{eq:Ujump}
\end{align}
For the momentum equation, the interfacial condition is the jump in the pressure across the interface given by
\begin{align}
  \left[p\right]_\Gamma=-\sigma\kappa-{\dot m}^2\left[\frac{1}{\rho}\right]_\Gamma\label{eq:Pjump}.
\end{align}
The term \(\sigma\kappa\) is the pressure caused by surface tension, where \(\sigma\) is the surface tension coefficient and \(\kappa=\nabla\cdot\bm n_\Gamma\) is twice the mean curvature.

The interfacial conditions for the remaining equations are more complex. The gradients of the vapor mass fraction and temperature are related implicitly through the definition of \(\dot m\). The value of \(\dot m\) varies locally on the interface, and its value reflects the local imbalance in thermal energy and species mass across the interface, 
\begin{align}
  \dot m = \frac{\left[\bm n_\Gamma\cdot k\nabla T\right]_\Gamma}{\left[h\right]},\label{eq:mdot_T}
\end{align}
and,
\begin{align}
  \dot m = \frac{\bm n_\Gamma\cdot\rho_GD\nabla Y}{Y^\Gamma -1},\label{eq:mdot_Y}
\end{align}
which are derived form \cref{eq:energy,eq:vapor_frac}, respectively. Since both \cref{eq:mdot_T,eq:mdot_Y} must be satisfied simultaneously, it follows that the gradients of vapor mass fraction and temperature are not independent at the interface. In \cref{eq:mdot_T}, the specific enthalpy, \(h\), in each phase is approximated as \(h=h_{sat}+C_p\left(T-T_{sat}\right)\). This leads to \(\left[h\right]=L_V+\left[C_p\right]\left(T_\Gamma-T_{sat}\right)\), where \(L_V\) is the specific latent heat of boiling. Accordingly, \(\left[h\right]\) is an effective latent heat, which varies as a function of temperature. Because \cref{eq:mdot_T} is derived from the thermal energy equation, \cref{eq:energy}, it implicitly presumes that thermal energy transfer at the interface dominates mechanical energy transfer. However, this presumption is consistent with the thermodynamic assumptions used in the work.

The relationship between \(T\) and \(Y\) is further complicated by the requirement of the thermodynamic equilibrium at the interface between the fuel and its vapor. This implies a functional relationship between the vapor mass fraction and the temperature fields. This relationship can be described by the Clausius-Clapeyron relations,
\begin{subequations}
\begin{align}
  X &=\exp\left(-\frac{L_VW_V}{R}\left(\frac{1}{T_\Gamma}-\frac{1}{T_{sat}}\right)\right), \text{ where} \label{eq:CCx}\\
  Y &= \frac{XW_V}{XW_V+(1-X)W_A}. \label{eq:CCy}
\end{align} \label{eq:CC}
\end{subequations}
Here \(X\) is the vapor mole fraction and \(W_V\) and \(W_A\) are the vapor and ambient gas molar masses, respectively. \Cref{eq:CC} is derived assuming the gas behaves ideally, hence the use of the gas constant, \(R\). The saturation temperature, \(T_{sat}\) are assumed to be known and are fixed in time and space. The temperature across the interface is assumed to be continuous, i.e., \(\left.T_L\right|_\Gamma=\left.T_G\right|_\Gamma=T_\Gamma\). \Cref{eq:mdot_T,eq:mdot_Y,eq:CCx,eq:CCy} must be satisfied simultaneously to correctly resolve the vaporization.

\section{Numerical Approach}\label{sec:numerics}
This section details the algorithms used to perform direct numerical simulations of vaporizing liquid-gas flows. The algorithms are embedded into NGA, a computational code for solving low-Mach, multiphase flows \cite{Desjardins2008b}. NGA uses a mixture of finite volume and finite difference schemes to solve the conservation equations (\cref{sec:physics}) on a staggered grid. Two time integration schemes are used which are detailed in \cref{sec:temps}. A detailed exposition of the methods used to solve the vaporization problem follows.

The numerical methods used in this work will be detailed in semi-discrete form.
For time evolving fields, superscripts are used. When necessary, secondary
superscripts are used to indicate sub-iterations, e.g., \(f^{n,k}\) occurs at
time level \(n\) at the \(k\) sub-iteration. By convention,
\(f^{n+1,0}=f^{n,m}\) where \(m\) is the total number of sub-iteration used in
each timestep. The elapsed time between time levels is represented as \(\Delta
t\). For spatially varying fields, subscripts are used to identify the position,
e.g., \(f_{i,j,k}^n\equiv f(\bm x,t)\), where \(\bm x\) is the centroid of the
grid cell indexed with \((i,j,k)\) and \(t=n\Delta t\). The distance between
the grid cell centroids \((i,j,k)\) and \((i+1,j,k)\) is given by \(\Delta x\). The
distances \(\Delta y\) and \(\Delta z\) are defined analogously. For brevity, superscript
and subscript notations will be omitted when the reference is unambiguous.

\subsection{Sharp Representation of Liquid-Gas Flows}
Before discussing the numerical methods themselves, it is worthwhile to discuss the strategy used to represent physical quantities and field variables in each grid cell. Following, Anumolu and Trujillo \cite{Anumolu2018a}, a strategy is said to be sharp when
\begin{enumerate}[nosep]
\item \label{enum:1} the interface position is known,
\item \label{enum:2} the thermophysical properties at the interface are discontinuous, and
\item \label{enum:3} fluxes are evaluated using only information from one phase.
\end{enumerate}
\Cref{enum:1} requires the use of an interface tracking scheme with explicit interface reconstruction. Such a scheme is used in this work (detailed in \cref{sec:transport}). With precise knowledge of the interface position, \cref{enum:2,enum:3} reduce to a statement on how information is averaged between the two phases. This work uses a mixture of both sharp and non-sharp representations.

A two-phase momentum equation is formed by taking the volume average of the Navier-Stokes equations, \cref{eq:NS}, evaluated in each phase. Using \(\alpha\), the effective density and viscosity in a computational cell can be defined as
\begin{align}
  \rho =\rho_L\alpha +\rho_G(1-\alpha ),\label{eq:density}
\end{align}
and
\begin{align}
  \mu =\mu_L\alpha+\mu_G(1-\alpha ).\label{eq:viscosity}  
\end{align}
The form of the momentum equations remains the same, simply replacing the effective density and viscosity for their single phase counterparts. A two-phase continuity equation is constructed similarly. Throughout the remainder of this work, the effective density and viscosity will be referred to using \(\rho\) and \(\mu\) while the phase-specific values will always use subscripts \(L\) and \(G\). The resulting representation of \(\rho\), \(\mu\), and \(\bm u\) is not sharp, since values from both sides of the interface are used in the computation of these quantities. Even so, this approach is preferred as it is consistent with the finite volume methodology used in this work.

In contrast, the temperature is represented using two fields, \(T_L\) and \(T_G\), which are defined within the liquid and gas, respectively. This approach follows Ma and Bothe \cite{Ma2013} who demonstrated that such a representation produced more accurate temperature profiles for vaporization simulations. This representation of the temperature field is sharp, since it uses only information from one side of the interface. The vapor mass fraction field is defined only in the gas, hence it is also treated sharply. For grid cells that contain both liquid and gas, data are computed at the phase barycenters \(\bm x_L\) and \(\bm x_G\), respectively. For example, in the cell \((i)\) the value of \(T_L\) is computed at \(\bm x_{L,i}\) while the value of \(T_G\) is computed at \(\bm x_{G,i}\).

\subsection{Time Integration}\label{sec:temps}
In general, the transport equations used in this work can be rewritten for a variable \(\phi\) in the form
\begin{align}
  \frac{\partial\left(\rho\phi\right)}{\partial t}=-\nabla\cdot\left(\rho\phi\bm u\right)+\nabla\cdot\left(\rho \xi\nabla\phi\right)+S(\bm x),\label{eq:timeInt}
\end{align}
where \(\xi\) is some appropriately defined diffusion coefficient and \(S\) represents sources to the equation. Selecting a time integration involves finding appropriate temporal discretization of these operators.

Two time integration schemes are used in this work. The default scheme is an iterative semi-implicit scheme given in \cite{Pierce2001} which is second order accurate. This time integration scheme corresponds to solving the semi-discrete equation
\begin{align}
  \frac{\left(\rho\phi\right)^{n+1,k+1}-\left(\rho\phi\right)^n}{\Delta t}&=-\nabla\cdot\left(\rho\phi\bm u\right)^{n+1/2,k}+\nabla\cdot\left(\rho \xi\nabla\phi\right)^{n+1/2,k}+S^{n+1/2,k}, \\
  \phi^{n+1/2,k}&=\frac{1}{2}\left(\phi^{n+1,k}+\phi^n\right),
\end{align}
with an arbitrary discretization of the spatial operators. It can be proven that this scheme is linearly stable~\cite{Pierce2001}. The second discretization scheme used in this work is a
\begin{align}
  \frac{\left(\rho\phi\right)^{n+1,k+1}-\left(\rho\phi\right)^n}{\Delta t}&=-\nabla\cdot\left(\rho\phi\bm u\right)^{n+1/2,k}+\nabla\cdot\left(\rho \xi\nabla\phi\right)^{n+1,k+1}+S^{n+1/2,k}, \label{eq:timeMonotone}\\
  \phi^{n+1/2,k}&=\frac{1}{2}\left(\phi^{n+1,k}+\phi^n\right).
\end{align}
The major difference between the two scheme is the time discretization of the diffusive fluxes, which are handled fully implicitly in the later expression. The reason for this is as follows. Consider a simplified transport equation consisting only of the diffusive component of the original equation, 
\begin{align}
  \frac{\partial\phi}{\partial t}=\frac{\partial}{\partial x}\left(\xi\frac{\partial \phi}{\partial x}\right),\label{eq:timeEase}
\end{align}
which has been taken in one dimension, for ease of exposition. (\Cref{eq:timeEase} exploits the constant density assumption by dividing density from both sides of the equation.) When discretized temporally using \cref{eq:timeMonotone} and spatially using the standard second order finite volume discretization
\begin{align}
  \frac{{\partial}^2\phi}{\partial x^2}=\frac{\phi_{i+1}-2\phi_i+\phi_{i-1}}{\Delta x^2},
\end{align}
this yields,
\begin{align}
  -Cu_{i+1}^{n+1,k+1}+(1+2C)u_i^{n+1,k+1}-Cu_{i-1}^{n+1,k+1}=u_i^n,\label{eq:monotone_discrete}
\end{align}
where \(C=\xi\Delta t/{\Delta x}^2\). This equation corresponds to a monotone scheme (see the proof in \cref{sec:monotoneProof}) and has the property of being not only linearly stable, but also non-oscillatory. Since vaporization is controlled by diffusive mechanisms via \cref{eq:mdot_T,eq:mdot_Y}, the non-oscillatory solution of diffusion is important. This idea is further explored in \cref{sec:stability}.

\subsection{Momentum Equation}
\begin{figure}
  \centering
  \includetikz{0.5\textwidth}{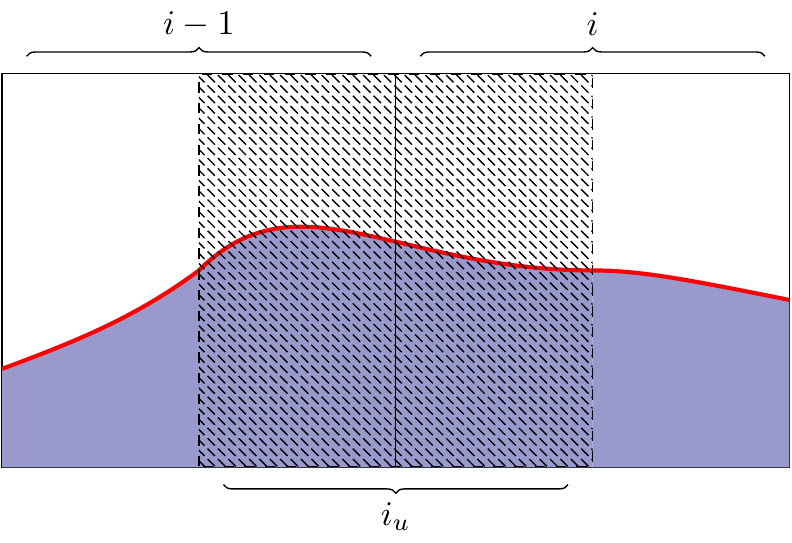}
  \caption{Construction of volume fraction on velocity cell, \(i_u\). Data from cells \(i-1\) and \(i\) are used to construct \(i_u\).}\label{fig:offset}
\end{figure}

The momentum equation is solved using a mass-momentum consistent advection strategy \cite{Desjardins2013}. The scheme is known to be stable for flows with large density ratios such as those encountered in a fuel-air system. The basic idea of the scheme is to update both \(\rho\) and \(\rho\bm u\) in time such that their discrete solution implies the discrete transport equation for \(\bm u\), via \(\bm u^{n+1}=\left(\rho \bm u\right)^{n+1}/\rho^{n+1}\). A overview of the solution is shown below for a one dimensional system.  Due to the staggered grid approach, velocities are solved on a mesh that is offset from the primary mesh (\cref{fig:offset}). A new set of effective densities and viscosities, \(\rho_u\) and \(\mu_u\), must be defined using \(\alpha_u\), the local fraction of liquid volume in the \(u\) velocity cell. This local volume fraction is computed by constructing the \(u\) velocity cell, and computing the liquid volume contained therein (\cref{fig:offset}). An interface reconstruction scheme must be used to compute the exact volume inside of the \(u\) velocity cell. The scheme used in this work is discussed in \cref{sec:transport}. At each time step, \(\alpha^n_u\) is computed from the cell centered field, \(\alpha^n\) at the current time level. The resulting algorithm to update \(u\) is
\begin{subequations}
  \begin{align}
    u^{n+1/2,k} &=\frac{1}{2}\left(u^{n+1,k}+u^n\right), \\
    %% u^{n+1/2} &=\frac{1}{2}\left(u^{n+1,k+1}+u^n\right), \\
    \frac{\rho_u^{n+1,k+1}-\rho_u^n}{\Delta t}+\frac{\partial\left(\rho_u^nu^{n+1/2,k}\right)}{\partial x}&=0, \label{eq:Mconst3} \\
    \frac{\rho_u^{n+1,k+1}u^{n+1,k+1}-\rho_u^nu^n}{\Delta t}+\frac{\partial\left(\rho_u^nu^nu^{n+1/2,k}\right)}{\partial x}&=RHS, \label{eq:Mconst4}
    %% S = 2\frac{\partial u^{n+1/2}}{\partial x}-\frac{2}{3}\frac{\partial u^{n+1/2}}{\partial x}
  \end{align}\label{eq:NStrue}
\end{subequations}
where \(RHS\) represents all non-advective terms in the Navier-Stokes equations. First \cref{eq:Mconst3} is solved for \(\rho^{n+1,k+1}_u\), and that value is used in \cref{eq:Mconst4} to update the velocity. The key aspect of this treatment is that the same discrete representation of the flux operator \(\partial /{\partial x}\) is used for advection in \cref{eq:Mconst3,eq:Mconst4}. Although any discretization may be used for this operator in general, care must be taken to ensure the stability of the scheme. In this work the flux operator is discretized using first order upwind in interface containing cells, and a second order finite volume discretization\cite{Desjardins2008b} otherwise. The specifics of the discretization including the definition of \(RHS\) are given in \cite{Desjardins2013}.

\subsection{Pressure Poisson Equation}
For low-Mach flows, the continuity equation is enforced through the pressure. In this work, Chorin splitting \cite{Chorin1968} is used to couple the momentum and continuity equations. First, an intermediate velocity is computed by solving the Navier-Stokes equations, \cref{eq:NS}, with zero pressure. This field is referred to as \({\bm u}^{NS}\), the Navier-Stokes velocity. Next the velocity field is updated with a pressure projection according to
\begin{align}
  {\bm u}^{n+1}={\bm u}^{NS}-\frac{\Delta t}{\rho}\nabla p.\label{eq:pressureSplit}
\end{align}
Taking the divergence of \cref{eq:pressureSplit} yields the Poisson equation for pressure,
\begin{align}
  \nabla\cdot\left(\frac{1}{\rho}\nabla p\right)=-\frac{1}{\Delta t}\nabla\cdot\left({\bm u}^{n+1}-{\bm u}^{NS}\right).\label{eq:PPE}
\end{align}
Conceptually, \Cref{eq:pressureSplit,eq:PPE} are similar to the projection technique used in purely incompressible flows. In that context, the velocity is projected onto its divergence-free part using a Helmholtz decomposition. However, in this work, it is not assumed that \(\bm u^{n+1}\) is incompressible, so the right hand side of \cref{eq:PPE} is nontrivial. By rearranging \cref{eq:cont}, we see
\begin{align}
  \nabla\cdot\bm u=-\frac{1}{\rho}\frac{D\rho}{Dt},\label{eq:divU}
\end{align}
and this value is used in \cref{eq:PPE}. For brevity, the notation
\begin{align}
  H=\frac{1}{\rho}\frac{D\rho}{Dt},\label{eq:H}
\end{align}
is adopted.

Here it should be noted the abuse of notation used in writing operators such as \(\nabla\cdot\bm u\) in the preceding discussions. The flow solver solves the equations of motion in the finite volume sense, meaning that the flow variables are updated in terms of fluxes through control volumes. This approach is consistent with directly solving the integral form of the conservation equations. However, in this work flow variables are not necessarily differentiable, requiring a careful definition of the various operators. As such, operators must be understood in the following sense. For some vector, \(\bm f\), define the operator,
\begin{align}
  \text{div}\left(\bm f\right)=\lim\limits_{\int\limits_\Omega dV\rightarrow 0}\frac{\int\limits_{\partial\Omega}\bm n\cdot \bm f dS}{\int\limits_\Omega dV}\label{eq:preGOD}
\end{align}
which is the net flux per unit volume out of \(\Omega\). For sufficiently smooth vector fields this limit yields
\begin{align}
  \text{div}\left(\bm f\right)=\frac{\partial f_i}{\partial x_i}=\nabla\cdot\bm f,\label{eq:GOD}
\end{align}
which follows from the Gauss-Ostrogradsky (Divergence) Theorem. In this work, all flow variables are continuous except at the interface, so relation \cref{eq:GOD} applies almost everywhere. Throughout this work, the suggestive notation \(\nabla\cdot\bm f\) has been used with the meaning \(\text{div}\left(f\right)\).

The form of \cref{eq:GOD} as a surface integral divided by an volume integral suggests that the surface density, \(\delta_\Gamma\), may be important in evaluating the divergence operator, \(\text{div}\), for discontinuous vectors. This can be seen by directly computing \(\text{div}\left(\bm u\right)\) near the interface. Define a volume \(\Omega_\Gamma\) by extending a small distance from the interface in both directions (i.e. in directions of both \(\bm n_\Gamma\) and \(-\bm n_\Gamma\)). In such a volume \(\text{div}\left(\bm u\right)_\Gamma\) satisfies,
\begin{align}
\lim\limits_{\int\limits_{\Omega_\Gamma}dV\rightarrow 0}\frac{\int\limits_{\Gamma}\left[\bm n_\Gamma\cdot \bm u\right]_\Gamma dS}{\int\limits_{\Omega_\Gamma}dV}=\lim\limits_{\int\limits_{\Omega_\Gamma}dV\rightarrow 0}\frac{\int\limits_{\Gamma}\left[\bm n_\Gamma\cdot \bm u\right]_\Gamma \delta_\Gamma dV}{\int\limits_{\Omega_\Gamma}dV}=\left[\bm n_\Gamma\cdot \bm u\right]_\Gamma \delta_\Gamma.\label{eq:discreteDIV}
\end{align}
In deriving \cref{eq:discreteDIV} it was important to assume that there is no jump in tangential velocity, such that the net flux through \(\partial\Omega_{\Gamma}\) equals the net flux through \(\Gamma\).

The overall value of \(\text{div}\left(\bm u\right)\) in a discrete computational cell contains three components: a component from the fully gas portion of the cell, a component from the fully liquid component of the cell, and an interfacial component as detailed in \cref{eq:discreteDIV}. Using this information, and expanding \(\left[\bm n_\Gamma\cdot\bm u\right]_\Gamma\) using \cref{eq:divU}, the pressure Poisson equation can be written as
\begin{align}
  \nabla\cdot\left(\frac{1}{\rho}\nabla p\right)=\frac{1}{\Delta t}\left(H-\dot m\left[\frac{1}{\rho}\right]_\Gamma\delta_\Gamma+\nabla\cdot{\bm u}^{NS}\right),\label{eq:PPEtrue}
\end{align}
where \cref{eq:H} is applied separately in each phase, so that
\begin{align}
  H=\alpha H_L+(1-\alpha )H_G.
\end{align}
\(H\) can change due to variations in the chemical composition, temperature, and thermodynamic pressure of the fluid. As a first step, the current work explicitly neglects changes in density due to chemical composition and temperature. This simplifies the exposition of the system thermodynamics considerably. Consistent with the low-Mach assumption, the thermodynamic pressure is assumed to vary only temporally. This implies that each phase's value of \(H\) is also a function of time only. Finally, we explicitly assume that the liquid is fully incompressible, implying that \(H\) can be nonzero only in the gas. 

By assumption, nonzero \(H\) correspond to a build up of pressure in the computational domain. Since sources due to chemical composition and temperature are neglected, this build up is due entirely to the evaporative flow. In the presence of outflow conditions, this flow is able to freely exit the domain. Accordingly, vaporization does not cause any pressure buildup, and \(H=0\). For closed boundary conditions, such as periodic boundary conditions, the flow cannot exit the domain, and a gradual rise in pressure occurs. For low-Mach flows, a consistent choice for this source is
\begin{align}
  H_G= \frac{-1}{\int\limits_GdV}\int\limits_{L\cup G}\left(-\dot m\left[\frac{1}{\rho}\right]_\Gamma\delta_\Gamma+\nabla\cdot{\bm u}^{NS}\right)dV.\label{eq:dRHO}
\end{align}
This choice ensures that the compatibility condition of the pressure Poisson equation is satisfied, i.e,
\begin{align}
  \int\limits_{L\cup G}\nabla\cdot\left(\frac{1}{\rho}\nabla p\right)dV=\int\limits_{L\cup G}\frac{1}{\Delta t}\left(H-\dot m\left[\frac{1}{\rho}\right]_\Gamma\delta_\Gamma+\nabla\cdot{\bm u}^{NS}\right)dV=0.
\end{align}
Satisfaction of this condition is sufficient to guarantee existence and uniqueness of the pressure correction, \(\nabla p\)~\cite{Majda1985}.

The gas density is then updated as
\begin{align}
  {\rho_G}^{n+1}={\rho_G}^n(1+\Delta tH_G),
\end{align}
which can be considered to be a low-Mach correction to the Poisson equation that accounts for gas phase compressibility.

The pressure Poisson equation, \cref{eq:PPEtrue}, is solved using a technique based upon the Ghost Fluid Method \cite{Liu2000}. An illustrative discretization in one dimension is defined below. The pressure Laplacian operator is discretized as, 
\begin{align}
  \frac{\partial }{\partial x}\left(\frac{1}{\rho}\frac{\partial p}{\partial x}\right)\approx \frac{1}{\Delta x}\left(\frac{1}{\rho_{u,i+1}}\frac{p_{i+1}-p_i}{\Delta x}-\frac{1}{\rho_{u,i}}\frac{p_i-p_{i-1}}{\Delta x}\right),
\end{align}
where the assumption of constant grid spacing has been used. The velocity divergence is written as
\begin{align}
  \frac{\partial u}{\partial x}\approx\frac{u_{i+1}-u_i}{\Delta x}.
\end{align}
The singular source term involving \(\delta\) requires a special treatment. Here this term is discretized as
\begin{align}
  \dot m\left[\frac{1}{\rho}\right]_\Gamma\delta_\Gamma\approx Q\dot m_i\left[\frac{1}{\rho}\right]_\Gamma\frac{1}{\Delta x},
\end{align}
the value \(Q\) is one in interface containing cells and zero otherwise. The pressure jump, \cref{eq:Pjump} is discretized using the standard GFM technique. As a single pressure field is solved for the entire domain, 
this discretization is not sharp, in the sense of \cite{Anumolu2018a}. However, this approach was chosen to be consistent with the one field approach to the velocity field. Furthermore, in practice, this technique only distributes the pressure jump over a region of one or two grid cells around the interface.

\subsection{Scalar Transport Equation}\label{sec:scalar}
Transport of the temperature and vapor mass fraction
fields is handled sharply to avoid artificial smearing of the profiles. One of the necessary aspects of a sharp
representation is that data for one phase are not computed using values from the
other phase. For this purpose, several auxiliary variables are defined that will
be necessary to discuss in the transport equations. At each time step, a gas volume fraction \(\alpha_G^n=1-\alpha^n\) and secondary liquid volume fraction \(\alpha_L^n=\alpha^n\) are constructed. To aid in the advection of scalars, a continuous velocity field is constructed for both the liquid and gas fields, labeled as \({\bm u}_L\) and \({\bm u}_G\), respectively. The fields are initialized using the true velocity field at the latest time level, \({\bm u}^{n+1,k}\), and the data are extrapolated using the zero normal gradient technique of Aslam \cite{Aslam2004}. In the technique, a PDE is solved in pseudotime, \(t_{ps}\), until steady state, and the steady state solution verifies the zero normal gradient property. The equation to extend an arbitrary field, \(\phi\), from the liquid into the gas can be expressed as
\begin{align}
  \frac{\partial \phi}{\partial t_{ps}}+A\bm n_\Gamma\cdot\nabla\phi=0\label{eq:PDEex}
\end{align}
where
\begin{align}
  A=\left\{
  \begin{array}{cc}
    0 & \text{if }\alpha_L=1,\\
    1 & \text{otherwise.}
  \end{array}
  \right.
\end{align}
A similar process is used to extrapolate from gas to liquid. The technique results in an extrapolated field that is constant along the direction of \(\bm n_\Gamma\). In \cite{Aslam2004}, a general technique for extrapolation to arbitrary polynomial order is discussed. However, for the current work it was found by experience that constant extrapolation yields superior numerical stability.

The transport of scalars uses a variation of the mass consistent scheme \cite{Desjardins2013}, outlined below for the vapor mass fraction
\begin{subequations}
  \begin{align}
    Y^{n+1/2,k} &=\frac{1}{2}\left(Y^{n+1,k}+Y^{n}\right) \\
    \frac{\rho_G^{n+1}\alpha_G^{n+1,k+1}-\rho_G^{n+1}\alpha_G^n}{\Delta t}+\frac{\partial\left(\rho_G^{n+1}\alpha_G^n{\bm u}_G\right)}{\partial x}&=0 \label{eq:Ytrue2}\\
    \frac{\rho_G^{n+1}\alpha_G^{n+1,k+1}Y^{n+1,k}-\rho_G^{n+1}\alpha_G^nY^n}{\Delta t}+\frac{\partial\left(\rho_G^{n+1}\alpha_G^nY^n{\bm u}_G\right)}{\partial x}&=
    \frac{\partial}{\partial x}\left(D\frac{\partial Y^{n+1/2,k}}{\partial x}\right)\label{eq:Ytrue3}
  \end{align}\label{eq:Ytrue}
\end{subequations}
First, \cref{eq:Ytrue2} is solved for \(\rho_G^{n+1}\alpha_G^{n+1,k+1}\) and then this value is used in \cref{eq:Ytrue3} to update \(Y\). For advection, the BQUICK algorithm \cite{Herrmann2006} is used. The algorithm modifies the QUICK advection routine, by locally switching to an simple upwind scheme wherever the computed \(Y^{n+1}\) becomes unbounded. For the present work, a second switching condition is used. Whenever the QUICK stencil crosses the interface, the simple upwind scheme is used. To use the BQUICK algorithm, reasonable bounds must be chosen for each scalar. For the mass fraction field, the maximum and minimum values of \(1\) and \(0\), respectively, are chosen. For the temperature fields, the maximum of the initial domain temperature, \(T_M\), and \(0\) are chosen. This choice is appropriate if there are no sources of heat, in which case \(T_M\) will always be the highest domain temperature physically possible. The overall accuracy of this scheme is first order, because the simple upwind scheme is first order accurate. In practice, higher accuracy has been observed \cite{Herrmann2006}.

\begin{figure}
  \centering
  \includetikz{0.618\textwidth}{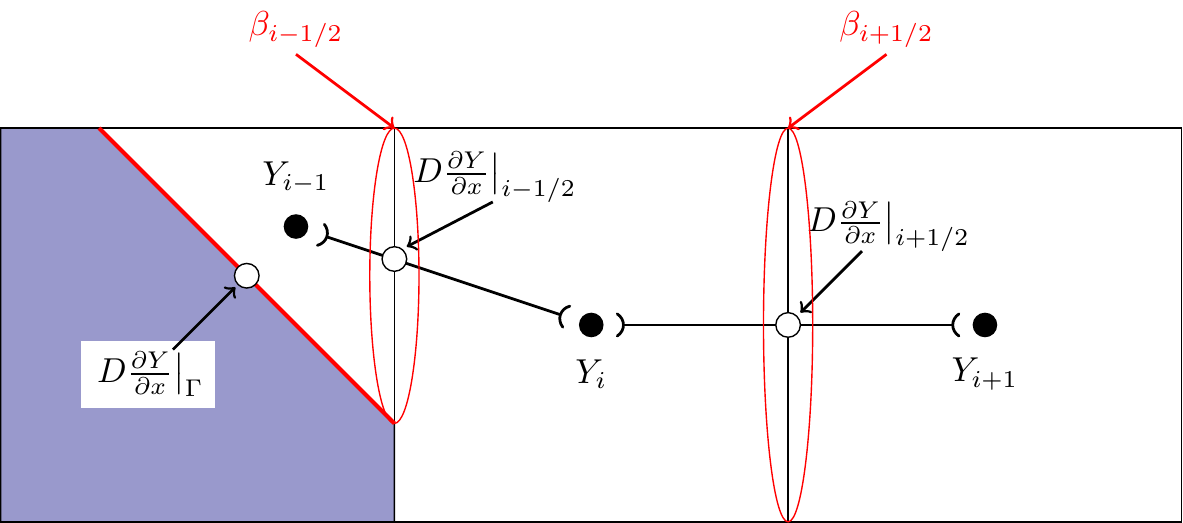}
  \caption{Diffusion fluxes of vapor mass fraction near the interface. Diffusion fluxes are computed on grid cell faces (white circles) using data from the gas phase barycenters (black circles).  Fluxes between cells are restricted to the gaseous fraction of the cell faces (circled in red) and the interface (thick red line).}\label{fig:diff}
\end{figure}

The treatment of diffusion in interface-containing cells requires special care. \Cref{fig:diff} demonstrates a region near the interface in one dimension. The interface does not occur inside either of cells \(i\) and \(i+1\), so the flux between the cells is computed using the second order finite volume discretization,
\begin{align}
  \frac{1}{2}(D_{i+1}+D_i)\frac{Y_{i+1}-Y_i}{\Delta x}.\label{eq:DiffSTD}
\end{align}
A different treatment is used to compute the flux whenever one (or both) of the two cells contains the interface, such is the case with cells \(i\) and \(i-1\). In this case, the flux is computed using the relative distances between the appropriate phase barycenters. For example, the discretization of the flux for the vapor mass fraction is
\begin{align}
  \beta_{i-1/2} D_{i-1/2}\frac{x_i-x_{i-1}}{\|\bm x_{G,i}-\bm x_{G,i-1}\|}(Y_i-Y_{i-1}).  \label{eq:DiffBARY}
\end{align}
Here \(x_i\) is the \(x\) axis component of \(\bm x_{G,i}\). The value \(\beta_{i-1/2}\) is the fraction of gas area on the face between cells \(i\) and \(i-1\). The value \(D_{i-1/2}\) is linearly interpolated between \(D_i\) and \(D_{i-1}\). Note that the VOF scheme does not represent the interface smoothly between cells, and accordingly, the gas area fraction may be different on either side of a cell face. In such a case, \(\beta_{i-1/2}\) is taken to be the minimum of the values.

In cell \(i-1\), an additional diffusive flux from the interface into the cell must be computed which satisfies
\begin{align}
  \int\limits_{\Gamma\cup\Omega_{i-1}} \bm n_\Gamma\cdot D\nabla Y dS,
\end{align}
or in terms of the surface area density,
\begin{align}
  \int\limits_{\Omega_{i-1}}\left(\bm n_\Gamma\cdot D\nabla Y\right)_\Gamma\delta_\Gamma dV.
\end{align}
The evaluation of the flux at the interface is the focus of \cref{sec:flux}. Although not detailed here \(T_G\) and \(T_L\) are solved in a manner analogous to the vapor mass fraction field.
\subsection{Thermodynamic Equilibrium at the Interface}\label{sec:flux}
Vaporization involves the flux of mass, momentum, and energy across the
interface. The imbalance of these fluxes is directly tied to the definition of
\(\dot m\) in \cref{eq:Pjump,eq:Ujump,eq:mdot_T,eq:mdot_Y}. This relation is
further complicated by the requirement of the system to satisfy thermodynamic equilibrium, i.e., \cref{eq:CC}. Because of this complexity, the solution of this nonlinear system is often neglected in the literature by assuming \textit{a priori} a limiting vaporization regime such as boiling conditions. There are, however, a few strategies in the literature capable of solving the general problem.

The first strategy used in this work is similar to a technique used in Ma and Bothe \cite{Ma2013}. In their work, they assume an error function profile for the temperature field, and substitute the thermodynamic relations to yield a nonlinear system that can be solved for the interface temperature. Such an assumption is appropriate for low \(\dot m\) vaporization (see \cref{sec:SPconv}), but fails to be appropriate for situations with appreciable convection. Instead, the current work constructs the system directly from the discretization of the governing equations. Recalling that gas and liquid temperatures (as well as mass fraction) are stored in their respective phase barycenter, specialized computational stencils must be use to accurately compute the gradient if the stencil includes the interface. Although several discretizations are possible, a general form of such a stencil is
\begin{align}
  \bm n_\Gamma\cdot\nabla T_G\approx \sum_i w_{G,i} T_{G,i},
\end{align}
where the summation is taken over computational cells near the interface and is restricted to the gas side. A liquid stencil can be defined similarly. Substituting these expressions into \cref{eq:mdot_T} yields
\begin{subequations}
  \begin{align}
    \dot m\left[h\right] = \sum_i w_{G,i}T_{G,i}+\sum_j w_{L,j} T_{L,j}+(w_{L,\Gamma}+w_{G,\Gamma })T_\Gamma,\text{ or} \\
    T_\Gamma= \frac{\dot m\left[h\right] -\left(\sum_i w_{G,i}T_{G,i}+\sum_j w_{L,j} T_{L,j}\right)}{w_{L,\Gamma}+w_{G,\Gamma }},\label{eq:stencil}
  \end{align}
\end{subequations}
where \(w_{G,\Gamma}\) and \(w_{L,\Gamma}\) refer to the gas-side and liquid-side coefficient to the interface-containing cell. Notably, knowledge of \(\dot m\) and the current temperature field allows the expression of the interface temperature in the form \(T_\Gamma={\mathcal T}(\dot m)\).

The system is closed by noting that discretizing \cref{eq:mdot_Y} yields a relationship of the form \(\dot m=\dot{\mathcal M}(Y_\Gamma)\), and the Clausius-Clapeyron relation, \cref{eq:CC}, requires \(Y_\Gamma=\mathcal{Y}(T_\Gamma)\). The entire system may thus be written as
\begin{align}
  T_\Gamma-{\mathcal T}({\dot {\mathcal M}}(\mathcal{Y}(T_\Gamma)))=0.\label{eq:fixed}
\end{align}
The solution of \Cref{eq:fixed} can be solved numerically using any root finding routine. A simple regula falsi solver was used in this work. The method outlined above solves the system with discrete consistency in the sense that each equation is satisfied simultaneously in each grid cell. This method for computing interface equilibrium will be henceforth referred to as MB method.

Rueda-Villegas et al.~\cite{Villegas2016} forgo the solution of the nonlinear problem, \cref{eq:fixed}. Instead, they use the approximation \(Y_\Gamma\approx Y(\bm x)\), and then invert the Clausius-Clapeyron relation to find \(T_\Gamma\). This method is not discretely consistent. To correct for this, they do not directly compute \(\bm n_\Gamma\cdot\nabla Y\) on the interface and instead substitute \(\frac{1-Y_\Gamma}{\rho_GD}\dot m\) where this quantity is needed. Note that the temperature field is used to compute \(\dot m\) through \Cref{eq:mdot_T}. This method is abbreviated as the REA method. The use of the local value of \(Y\) for the interface value introduces a first order error into the system. In the current work, we introduce a modified version of the REA technique that instead extrapolates the \(Y\) field to the interface with second order accuracy. The importance of this higher order accuracy will be discussed in \cref{sec:interface}. This modified technique is referred to as REA2.

In this work, two stencils are used to discretize the vaporization fluxes at the
interface. The pressure Poisson equation, \cref{eq:PPE}, and the VOF transport
equation, \cref{eq:VOF}, use \(\dot m\) but have no direct knowledge of the
scalar fields. The scalar transport equations, in turn, are effectively independent of each other, only being coupled implicitly through the boundary conditions, \(T_\Gamma\) and \(Y_\Gamma\). This leads to the following splitting. Inside of scalar transport equations, the vaporization fluxes are implemented using a first order accurate gradient approximation. For the gas temperature, this takes the form,
\begin{align}
  \bm n_\Gamma\cdot\nabla T_G=\frac{T_G^{n+1,k+1}-T^{n+1}_{G,\Gamma}}{\|\bm x_G-\bm x_\Gamma\|},\label{eq:upwind}
\end{align}
where \(\bm x_G\) and \(\bm x_\Gamma\) are the gas phase barycenter and interface barycenter within the computational cell. Note that \cref{eq:upwind} is discretized using the \(n+1\) time level, so that its use corresponds to a fully implicit time integration. Furthermore, this discretization is monotone and therefore does not introduce oscillations to the solution of the scalar fields. This discretization is important. Since the interface moves freely through the domain, \(\|\bm x_G-\bm x_\Gamma\|\) can approach zero making the interfacial flux term be of arbitrary magnitude. For interface containing cells, this term usually dominates the scalar transport phenomena, so its accurate resolution is paramount.

For the pressure Poisson equation and VOF transport equation it was found that a higher order treatment of \(\dot m\) was necessary to achieve a convergent solution. First, each scalar is extended across the interface using linear extrapolation \cite{Aslam2004}. Then a least squares fit to the temperature profile in each phase is computed over each cell and its \(5^{dim}-1\) neighbors, in \(dim=3\) dimensions. Gradients are calculated from the least square fit, and finally \cref{eq:mdot_T} is used to compute \(\dot m\). For example, in one dimension the cell \(\Omega_k\) is fit with the functional form
\begin{align}
  T=c_0+c_1(x-x_k),\label{eq:LSQ}
\end{align}
yielding the system
\begin{subequations}
\begin{align}
  \bm{\mathcal L}\bm c&=\bm T \\
  {\bm {\mathcal L}}_{i,j} &= \frac{1}{j!}{(x_{G,k+i}-x_k)}^j \\
  \bm c &= [c_0;c_1]^\top \\
  {\bm T}_i &= T_{G,k+i} \\
  i &\in \{-2,-1,0,1,2\} \\
  j &\in \{0,1\}
\end{align}
\end{subequations}
which is solved in the least squares sense, i.e., \(\bm c=({\bm {\mathcal L}}^\top{\bm {\mathcal L}})^{-1}{\bm {\mathcal L}}^\top\bm T\). It follows from \cref{eq:LSQ} that the coefficients \(c_i\) are the least squares fits for the temperature profile at the cell centroid. In practice, this technique does not lead to any issues with stability, due to the weak coupling of \cref{eq:PPE,eq:VOF} with the scalar transport equations. The technique provides another interesting property. As noted in \cite{Anumolu2018a}, knowledge of \(\dot m\) is needed not only in interface-containing cells, but also in a narrow band of cells around the interface. The treatment of \(\dot m\) given in our work provides a natural framework to extend \(\dot m\) across into such a band, namely, by computing \cref{eq:mdot_T} locally in each cell using the extended temperature field. %% This also requires some technique to extend the definition of the normals themselves, accross the interface.
\subsection{Interface Transport and Divergence-free Velocity Extrapolation}\label{sec:transport}
Interface transport is accomplished using a second order accurate, unsplit, geometric
volume of fluid method \cite{Owkes2014}. The volume fraction is transported using the
extrapolated liquid velocity, \({\bm u}_L\). The VOF method used in
this work ensures consistency between the discrete velocity divergence and the
flux volumes used in the transport step. An implication of this feature is that
errors in divergence of the extrapolated velocity field lead to erroneous variations in the liquid volume. To remove this source of error, the extrapolation is completed in two steps. First, the velocity is extrapolated using constant extrapolation \cite{Aslam2004}. The resulting velocity is labeled \({\bm u}_L^*\). Next, \({\bm u}_L^*\) is projected onto its divergence-free part using
%% \begin{align}
%%   \frac{\partial\alpha}{\partial t}+\nabla\cdot\left(\alpha\bm u\right)=-\frac{\dot m}{\rho_L}\delta_\Gamma.
%% \end{align}
\begin{align}
  {\bm u}_L^n={\bm u}_L^*-\nabla W,
\end{align}
where W is a potential derived from the Helmholtz-type equation,
\begin{align}
  aW+{\nabla}^2 W=\nabla\cdot{\bm u}_L^*.\label{eq:PPE_L}
\end{align}
The coefficient \(a\) is zero for all liquid-containing cells and cells that are within three grid cells of the interface, which reduces it to a Poisson equation in this region. For all other cells, \(a\) is set to an arbitrary constant, here given by \(\frac{1}{{\Delta t}^2}\). The advantage of solving \cref{eq:PPE_L} over a Poisson equation is that the given technique does not need to satisfy the compatibility relation
\begin{align}
  \int\limits_{L\cup G}\left(\nabla\cdot{{\bm u}_L^*}\right)dV=0,\label{eq:compat_uL}
\end{align}
\Cref{eq:compat_uL} will not hold, in general, for the extrapolated velocity field.

The volume of fluid advection equation, \cref{eq:VOF}, can be solved in a modified form. The interface recession term, \(\dot{m}\left/\rho_L\delta_\Gamma\right.\), is rewritten in terms of a recession velocity, \(\dot{m}/\rho_L\bm n_\Gamma\). Furthermore, since \(\nabla\cdot{\bm u}_L=0\) everywhere that \(\alpha\) is nonzero, the resulting equation simplifies to
\begin{align}
  \frac{\partial\alpha}{\partial t}+\left({\bm u}_L-\frac{\dot m}{\rho_L}\bm n_\Gamma\right)\cdot\nabla\alpha=0.
\end{align}
In practice, the velocity \(\left({\bm u}_L-\frac{\dot m}{\rho_L}\bm n_\Gamma\right)\) is used as the velocity seen by the VOF solver.

\begin{figure}
  \centering
  \includetikz{0.5\textwidth}{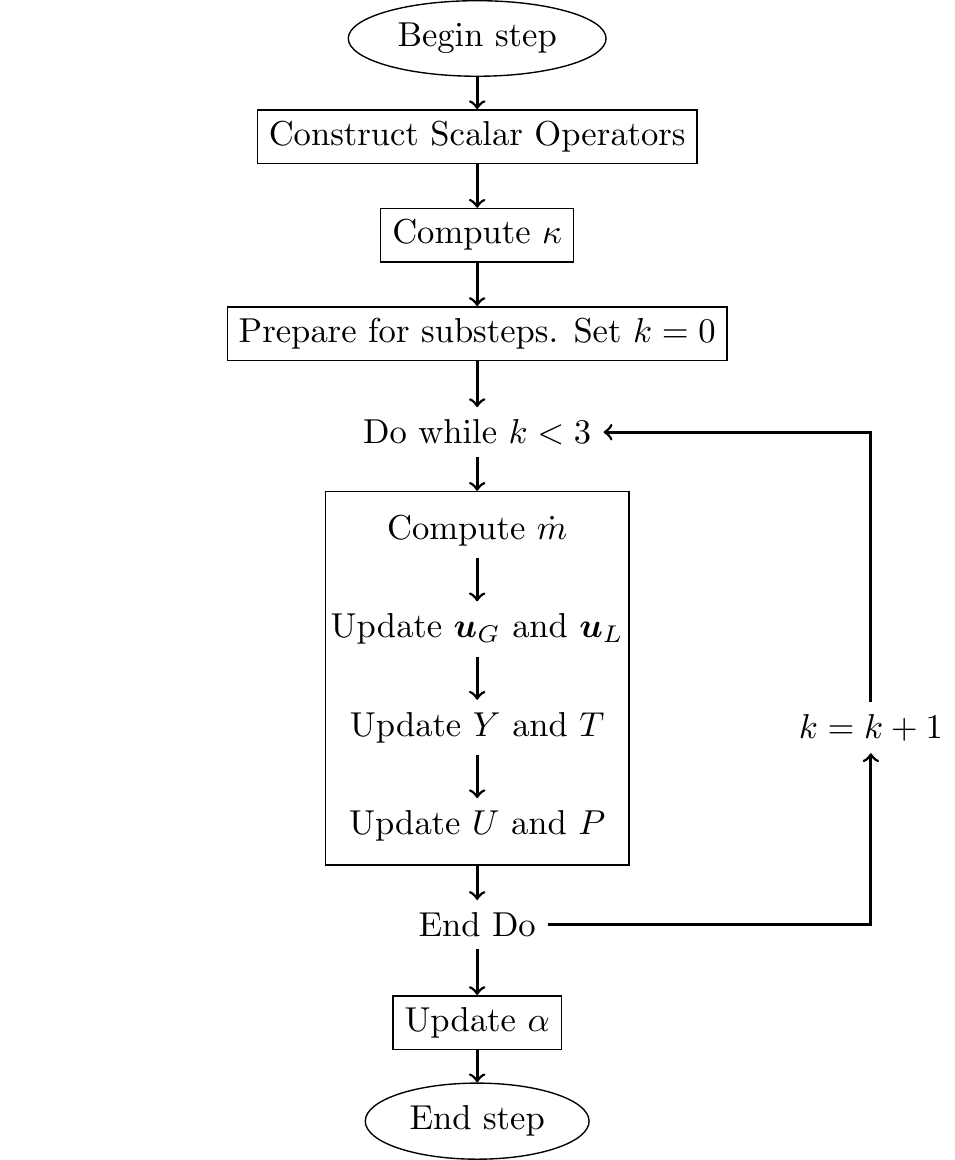}
  \caption{Outline of a single time step of the flow solver. Sub-iterations are indexed with \(k\).}\label{tikz:flow}
\end{figure}

The volume fraction field is used to compute all relevant geometric information for the solver. The interface is reconstructed from \(\alpha\) using the Piecewise Linear Interface Calculation (PLIC) method with ELVIRA \cite{Pilliod2004} which also defines the interface normal \(\bm n_\Gamma\). PLIC represents the interface in each cell as a simple polygon which separates a liquid polyhedron from a gas polyhedron. The representation of the geometry in terms of polytopes allows for a simple algorithm to compute the area (volume) of each polygon (polyhedron) by subdividing it into triangles (tetrahedra) \cite{Owkes2014}. The curvature \(\kappa\) is computed using the mesh decoupled height function approach \cite{Owkes2015}. A diagram showing the flow of a single time step of the solver is shown in \cref{tikz:flow}.
\section{Verification in One Dimension}\label{sec:results}
A suite of numerical tests is performed to study the accuracy of the code.
First, spatial convergence is demonstrated against a well-studied heat transfer
limited case. Second, this case is modified to study vaporization without the
constraint of heat transfer limitation. All simulations are
required to satisfy a CFL stability restraint, which is a numerical limitation
on the possible time step that can be used in a simulation. The CFL numbers depend on the physical problem being studied as well as the numerical discretization of the governing equation. The CFL numbers for advection, viscous diffusion, mass diffusion, temperature diffusion, and surface tension are
\begin{align}
  %% C_u=&\frac{\Delta t}{\Delta_m} \max\limits_{L\cup G} u_M\label{eq:cflu} \\
  C_u=&\frac{\Delta t}{\Delta_m} \max\limits_{L\cup G} u_M\label{eq:cflu} \\
  C_\mu =& \frac{4\mu\Delta t}{{\Delta_m}^2} \label{eq:cflm} \\
  C_D =& \frac{4D\Delta t}{{\Delta_m}^2} \label{eq:cflD} \\
  C_\lambda =& \frac{4\lambda\Delta t}{{\Delta_m}^2} \label{eq:cfll} \\
  C_\sigma =& \Delta t\sqrt{\frac{\sigma}{\left(\rho_L+\rho_G\right)\left(\Delta_m/{2\pi}\right)^{3}}}\label{eq:cfls}
\end{align}
where the notations \(u_M=\sum\limits_{i\in 1,2,3}|u_i|\)and \(\Delta_m =\min\left(\Delta x,\Delta y,\Delta z\right)\) are used.
The thermal diffusivity, \(\lambda =k/(\rho C_p)\), has been introduced for convenience. \Cref{eq:cflm,eq:cfll} should be understood to apply to both the liquid and the gas values of \(\mu\) and \(\lambda\). For the numerical schemes used in this work, the conditions \(C_\sigma <1\) and \(C_u <1\) must be satisfied, whereas the relations \cref{eq:cflm,eq:cflD,eq:cfll} are relaxed due to the use of implicit solvers.
\subsection{Spatial Convergence}\label{sec:SPconv}
\begin{figure}
  \includetikz{\textwidth}{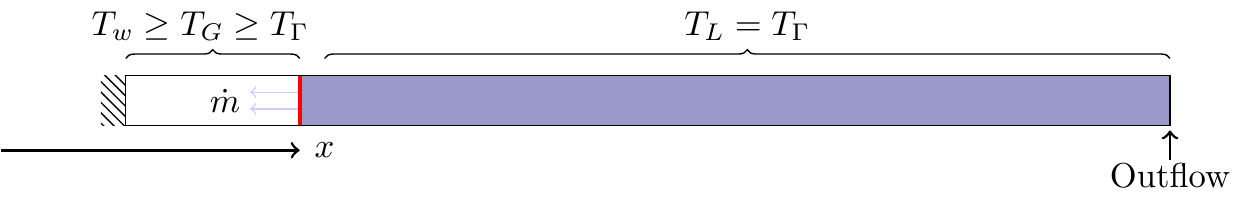}
  \includetikz{\textwidth}{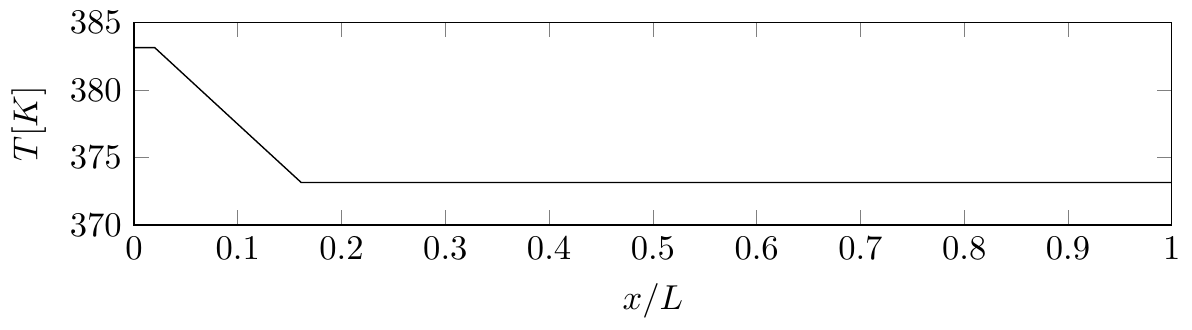}
  \caption{Geometry for Neumann's problem. The gas is heated on the left by a wall of constant temperature, \(T_w\). The liquid vaporizes into the gas, and is free to flow from the right boundary. Sample temperature profile shown.}\label{fig:Neumann}
\end{figure}

Due to the inherent complexities of the vaporizing problem, there are very few analytical solutions to verify against. One problem that does have such a solution is Neumann's problem, whose solution is given in the paper by Hardt and Wondra \cite{Hardt2008} The vaporization problem is solved in \textit{one dimension} with an assumption of a slow moving interface and negligible convective effects. A schematic is shown in \cref{fig:Neumann}. A region of hot gas heats the liquid, causing vaporization. Dirichlet conditions are specified for the temperature and mass fraction of the gas at the left boundary whereas the liquid is able to flow freely from the right boundary. The domain is \(1\) mm in length.

Under the assumptions given here, the energy equation, \cref{eq:energy}, reduces to the unsteady diffusion equation with time varying boundary. Define the initial and current interface location as \(x_0\) and \(x_\Gamma\), respectively. The solution is given in terms of the error function as
\begin{align}
    T = T_\infty+\frac{T_\Gamma-T_\infty}{\erf (\ell)}\erf\left(\frac{x-x_0}{2\sqrt{\lambda t}}\right). \label{eq:Neu_sol}
\end{align}
In \cref{eq:Neu_sol}, the interface position has been assumed to vary as
\(x_\Gamma = x_0+2\ell\sqrt{\lambda t}\). The diffusion layer value, \(\ell\), must be solved from thermodynamic variables as
\begin{align}
  \ell\exp (\ell^2)\erf (\ell)=\frac{C_{p,G}(T_\Gamma-T_\infty)}{\sqrt{\pi}L}.\label{eq:Neu_delta}
\end{align}
In their work, Hardt and Wondra solve only for the gas temperature field, however, the same analysis can be performed for the mass fraction field, yielding
\begin{align}
    Y = Y_\infty+\frac{Y_\Gamma-Y_\infty}{\erf (\ell_Y)}\erf\left(\frac{x-x_0}{2\sqrt{D t}}\right) \label{eq:Neu_solY}.
\end{align}
In general, an equation similar to equation \cref{eq:Neu_delta} can be written for \(\ell_Y\), however for the unity Lewis number case, the uniqueness of the interface position implies \(\ell_Y=\ell\). By assumption, the liquid temperature field remains at the interface temperature, \(T_\Gamma\). \Cref{eq:Neu_sol,eq:Neu_solY} can be combined with the condition of equality of the two expression for \(\dot m\), \cref{eq:mdot_T,eq:mdot_Y}, to yield an analytical relation between the interface temperature and mass fraction,
\begin{align}
  T_\Gamma=T_\infty+\frac{L}{C_p}\left(\frac{D}{\lambda}\right)^{1/2}\frac{Y_\Gamma-Y_\infty}{Y_\Gamma-1}.\label{eq:Neu_temp}
\end{align}
\Cref{eq:Neu_temp} should be combined with the Clausius-Clapeyron relation, \cref{eq:CC} to close the system.

The parameters used in this study mimic the properties of a vaporizing water-air system. They are listed in \cref{tab:Neumann}. The fluid and thermal properties listed are the same as those used in the work of Hardt and Wondra \cite{Hardt2008}. A unity Lewis number is imposed, and for consistency with their work, the vapor was given the molar mass of water. Two test cases are performed with these parameters by varying \(T_\infty\) and \(Y_\infty\).

\begin{table}
  \centering
  \begin{tabular}{l c l l}
    \toprule
    Property & Units & Gas & Liquid \\
    \midrule
    \(\rho\) & \(\text{kg}/\text{m}^3\) & \(1\) & \(1000\) \\
    \(\mu\) & \(\text{kg}/\left(\text{m}\cdot\text{s}\right)\) & \(1\times 10^{-5}\) & \(1\times 10^{-2}\) \\
    \(C_p\) & \(\text{J}/\left(\text{kg}\cdot\text{K}\right)\) & \(1000\) & \(1000\) \\
    \(k\) & \(\text{W}/\left(\text{m}\cdot\text{K}\right)\) & \(1\times 10^{-2}\) & \(1\times 10^{-1}\) \\
    \(D\) & \(\text{m}^2/\text{s}\) & \(1\times 10^{-5}\) & \textemdash\\
    \(M\) & \(\text{kg}/\text{mol}\) & \(0.018\) & \textemdash\\
    \(\sigma\) & \(\text{N}/\text{m}\) & \textemdash & \(0.01\) \\
    \(L_V\) & \(\text{J}/\text{kg}\) & \textemdash & \(1\times 10^6\) \\
    \(T_{boil}\) & \(\text{K}\) & \textemdash & \(373.15\) \\
    \bottomrule
  \end{tabular}
  \caption{Fluid properties for Neumann's problem.}\label{tab:Neumann}
\end{table}

\subsubsection*{Case 1: Film Boiling}
\begin{figure}
  %% \centering
  \begin{subfigure}{0.49\textwidth}
    \centering
    \includetikz{\linewidth}{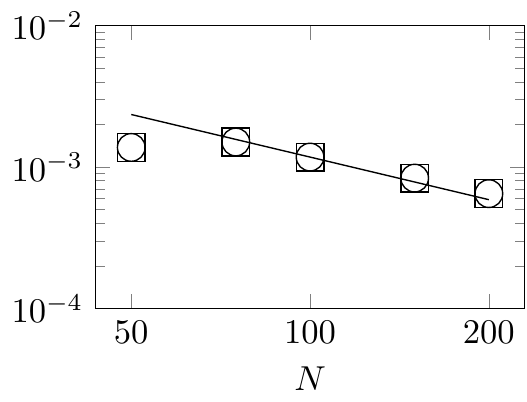}
    \caption{\(L_1\) error in gas temperature.}\label{fig:Neu1:L1}
  \end{subfigure}
  \begin{subfigure}{0.49\textwidth}
    \centering
    \includetikz{\linewidth}{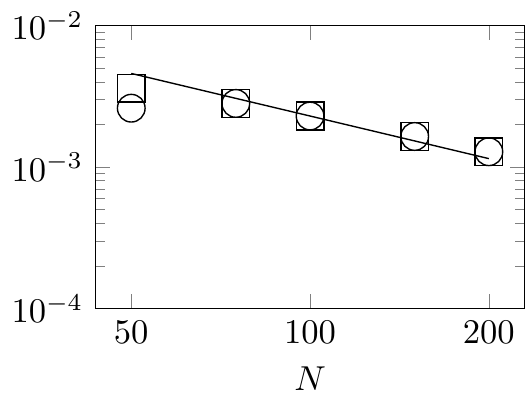}
    \caption{\(L_\infty\) error in gas temperature.}\label{fig:Neu1:Linf}
  \end{subfigure}
  \begin{subfigure}{0.49\textwidth}
    \centering
    \includetikz{\linewidth}{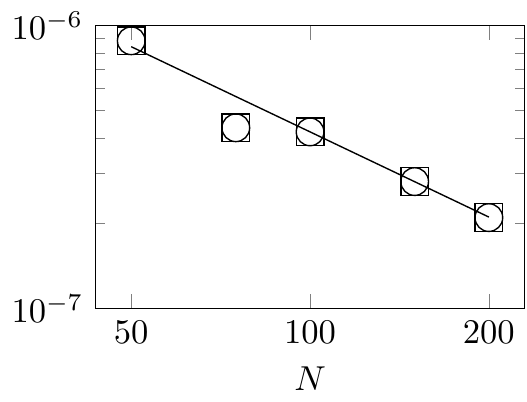}
    \caption{\(L_\infty\) error in liquid velocity.}\label{fig:Neu1:V_Linf}
  \end{subfigure}
  %%   \begin{subfigure}{0.49\textwidth}
  %%   \centering
  %%   \includetikz{\linewidth}{hardt_sol}
  %%   \caption{Gas temperature on 50 cell grid.}\label{fig:Neu1:sol}
  %% \end{subfigure}
  \caption{Case 1. Errors computed at \(t=0.1\text{ s}\). MB (\reftikz{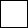}{tikz:HL1:MaBo}), REA (\reftikz{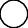}{tikz:HL1:REAL}). A first order trend line has been added.}\label{fig:Neu1}
\end{figure}

The first test, Case 1, corresponds exactly to that used in the paper of Hardt and Wondra,
which uses \(T_\infty=383.15\text{ K}\) and \(Y_\infty=1\), leading to
\(T_\Gamma=373.15\text{ K}\) and \(Y_\Gamma=1\). To aid in the imposition of the boundary
conditions, \(x_0\) is chosen to be inside the computational domain at
\(x_0=0.02L=2\times 10^{-5}\text{ m}\). We chose to initialize the simulation with the analytical
solution at time \(t=0.1\text{ s}\). Simulations are performed of vaporization using the
given parameters between times \(t=0.\text{ s}\) and \(t=0.2\text{ s}\). We initialize the
liquid field to the interface temperature, \(T_\Gamma\). Both the MB technique
and REA technique are used to resolve interface equilibrium. A grid convergence
study is performed using \(50\), \(75\), \(100\), \(150\), and \(200\)
grid cells in the domain. All simulations are performed for a fixed \(C_\lambda=C_D=2\). The analytical solution at \(t=0.3\text{ s}\) is shown in \cref{fig:Neumann} for the 50 cell grid. It can be seen that both the MB and REA simulations match well with the expected solution. We define error norms for the gas temperature as
\begin{align}
  \varepsilon & = \left|T-T_{exact}\right|, \\
  L_1 & = \int\limits_G \varepsilon dV\left/\int\limits_GdV\right., \text{ and} \label{eq:L1} \\
  L_\infty & = \max_G\varepsilon. \label{eq:Linf}
\end{align}
The vapor mass fraction and liquid temperature fields are
omitted, because they remain within machine precision of their initial values
for all cases, as expected. The \(L_1\) and \(L_\infty\) errors in gas temperature are plotted in
\cref{fig:Neu1:L1,fig:Neu1:Linf}. It can be seen that approximately first order convergence is noted, which is consistent with the overall scheme accuracy. Furthermore, the two schemes produce almost identical results.

\subsubsection*{Case 2: General Vaporization}\label{sec:interface}
While valuable as a comparison to previous work, the standard Neumann's problem  does not test the accuracy of the dynamic interface computation, in a situation with nontrivial mass fraction field. A second case, Case 2, was performed using the same grid and fluid parameters with \(T_\infty=323.15\text{ K}\) and \(Y_\infty=0.2\). This yields \(T_\Gamma=296.163\text{ K}\) and \(Y_\Gamma=0.221022\). The simulation was performed between times \(t=0.01\) and \(0.1\text{ s}\). A grid convergence study is performed with \(C_\lambda=C_D=2\). For a more complete analysis of this field, additional error measurements are taken. Errors for the vapor mass fraction, the liquid temperature, and liquid velocity are defined in manner analogously to \cref{eq:L1,eq:Linf}. The gas temperature and vapor mass fraction are referenced to the analytical solution, \cref{eq:Neu_sol,eq:Neu_solY}. The liquid temperature is referenced to initial value, which is predicted to be constant in time. The velocity error is computed relative to the following solution. Since the problem is one dimensional, the continuity equation predicts a piecewise constant velocity, with a discontinuity only at the interface. The liquid velocity can then be predicted from the definition of \(\dot m\)
\begin{align}
  \dot m=\rho_L(u_S-u_L)=\rho_G(u_S-u_G),\label{eq:mdot_Neumann}
\end{align}
where \(u_L=-\bm n_\Gamma\cdot\bm u_L\), \(u_G=-\bm n_\Gamma\cdot\bm u_G\), \(u_S=-\bm n_\Gamma\cdot\bm u_S\). Note that \cref{eq:mdot_Neumann} has been expressed such that positive speeds corresponds to motion in the positive \(x\) direction. This expression rearranges into,
\begin{align}
  u_L=\left(1-\frac{\rho_G}{\rho_L}\right)u_S+\frac{\rho_G}{\rho_L}u_G.
\end{align}
Since \(x_\Gamma = x_0+2\ell\sqrt{\lambda t}\), it follows
\begin{align}
  u_S=\ell\sqrt{\frac{\lambda}{t}}.\label{eq:Sdot}
\end{align}
With this information, the liquid velocity can be expressed as 
\begin{align}
  u_L=\ell\sqrt{\frac{\lambda}{t}}\left(1-\frac{\rho_G}{\rho_L}\right),\label{eq:NeuVelo}
\end{align}
where \(u_G=0\) has been substituted.

\begin{figure}
  %% \centering
  \begin{subfigure}{0.49\textwidth}
    \centering
    \includetikz{\linewidth}{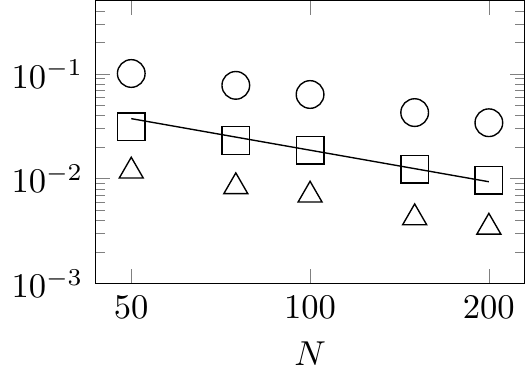}
    \caption{\(L_1\) error in gas temperature.}\label{fig:Neu2:L1}
  \end{subfigure}
  \begin{subfigure}{0.49\textwidth}
    \centering
    \includetikz{\linewidth}{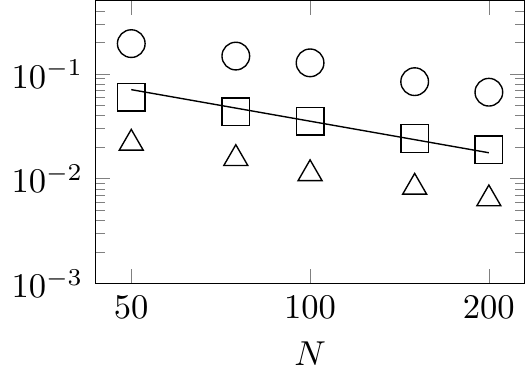}
    \caption{\(L_\infty\) error in gas temperature.}\label{fig:Neu2:Linf}
  \end{subfigure}
  \begin{subfigure}{0.49\textwidth}
    \centering
    \includetikz{\linewidth}{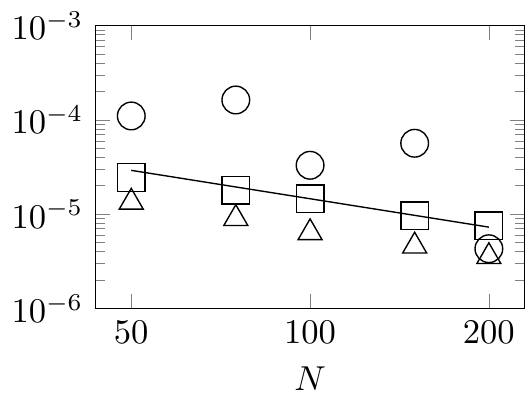}
    \caption{\(L_1\) error in vapor mass fraction.}\label{fig:Neu2:L1Vap}
  \end{subfigure}
  \begin{subfigure}{0.49\textwidth}
    \centering
    \includetikz{\linewidth}{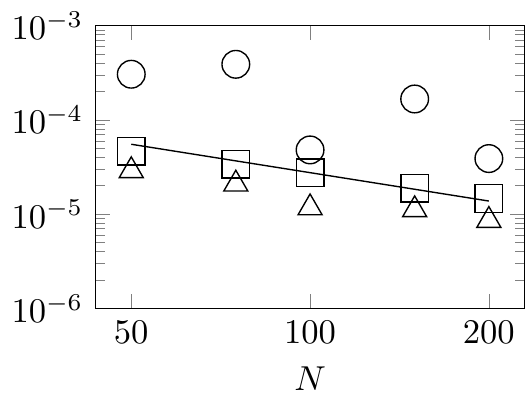}
    \caption{\(L_\infty\) error in vapor mass fraction.}\label{fig:Neu2:LinfVap}
  \end{subfigure}
  \begin{subfigure}{0.49\textwidth}
    \centering
    \includetikz{\linewidth}{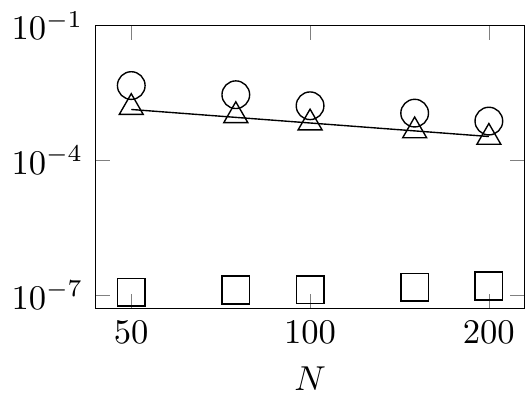}
    \caption{\(L_1\) error in liquid temperature.}\label{fig:Neu2:L1Liq}
  \end{subfigure}
  \begin{subfigure}{0.49\textwidth}
    \centering
    \includetikz{\linewidth}{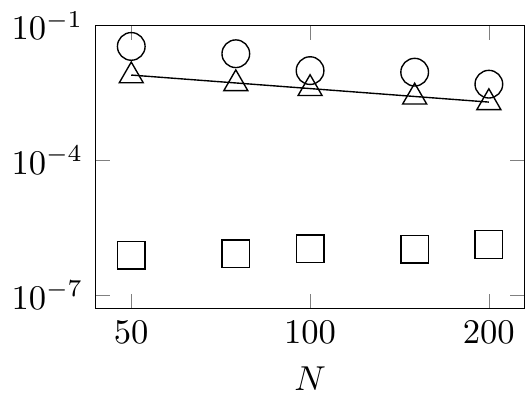}
    \caption{\(L_\infty\) error in liquid temperature.}\label{fig:Neu2:LinfLiq}
  \end{subfigure}
  \begin{subfigure}{0.49\textwidth}
    \centering
    \includetikz{\linewidth}{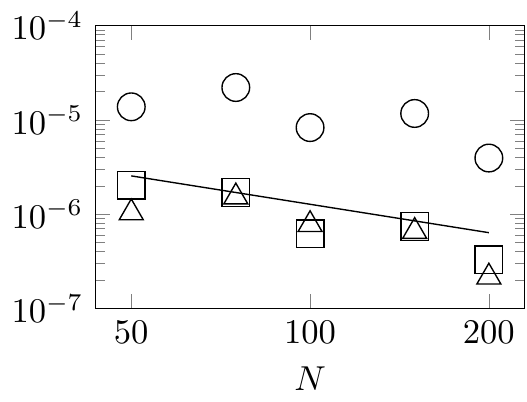}
    \caption{\(L_\infty\) error in liquid velocity.}\label{fig:Neu2:LinfVelo}
  \end{subfigure}
  \caption{Case 2. Errors computed at \(t=0.1\text{ s}\). MB (\reftikz{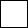}{tikz:MVelo_Linf:MaBo}), REA (\reftikz{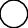}{tikz:MVelo_Linf:REAL}), REA2 (\reftikz{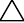}{tikz:MVelo_Linf:REAL2}). A first order trend line has been added.}\label{fig:Neu2}
\end{figure}

The gas temperature converges with first order accuracy in both \(L_1\) and \(L_\infty\) sense for all three techniques (\cref{fig:Neu2:L1,fig:Neu2:Linf}). However, it is interesting to note that errors using REA are consistently higher than either of the other two techniques, and errors using REA2 are consistently the lowest. A similar trend is seen for the vapor mass fraction field in \cref{fig:Neu2:L1Vap,fig:Neu2:LinfVap}, although greater variance about the trend is seen for the REA technique. The liquid temperature field (\cref{fig:Neu2:L1Liq,fig:Neu2:LinfLiq}) demonstrates a different trend. While first order convergence is seen using the REA and REA2 techniques, no convergence is seen using MB. Indeed, MB appears to divergence under grid refinement. However, the overall liquid temperature error associated with the MB technique is orders of magnitude lower than that of the REA or REA2 techniques. The liquid velocity error is demonstrated in \cref{fig:Neu2:LinfVelo}. Since the liquid velocity is spatially constant, there is no distinction between \(L_1\) and \(L_\infty\) errors. This property has been verified, but the redundant plot is not shown here. A trend of approximately linear convergence is displayed, although the trend appears to have some variation.

\section{Vaporization of a Curved Interface: Convergence and Stability}\label{sec:2D}
The solutions to the vaporization problems presented so far have been analytical solutions in one dimension. These solutions, while interesting for verification, do not allow for a more full test of the capabilities of the solver. One of the goals of this work was to develop a robust numerical solver for simulating vaporizing multiphase flows. However solutions in 1D cannot fully test this. Many of the most challenging aspects of multiphase flow simulation are due to non-alignment between the mesh and flow structures; however, in one dimension all flow structures are aligned with the mesh. To understand what influence, if any, mesh alignment has on the flow solver simulations are performed in two dimensions with a complex, curved interface. Since no analytical solution is known to this problem, errors are approximated by comparing simulation data to those of the same simulation run on a fine grid. This section also serves to test the numerical stability of the solver, as  a temporal convergence study is performed. Results are compared for both the semi-implicit and fully implicit numerical time integrator.

\begin{figure}
  \centering
  \begin{subfigure}{0.49\textwidth}
    \centering
    \includegraphics[width=\linewidth]{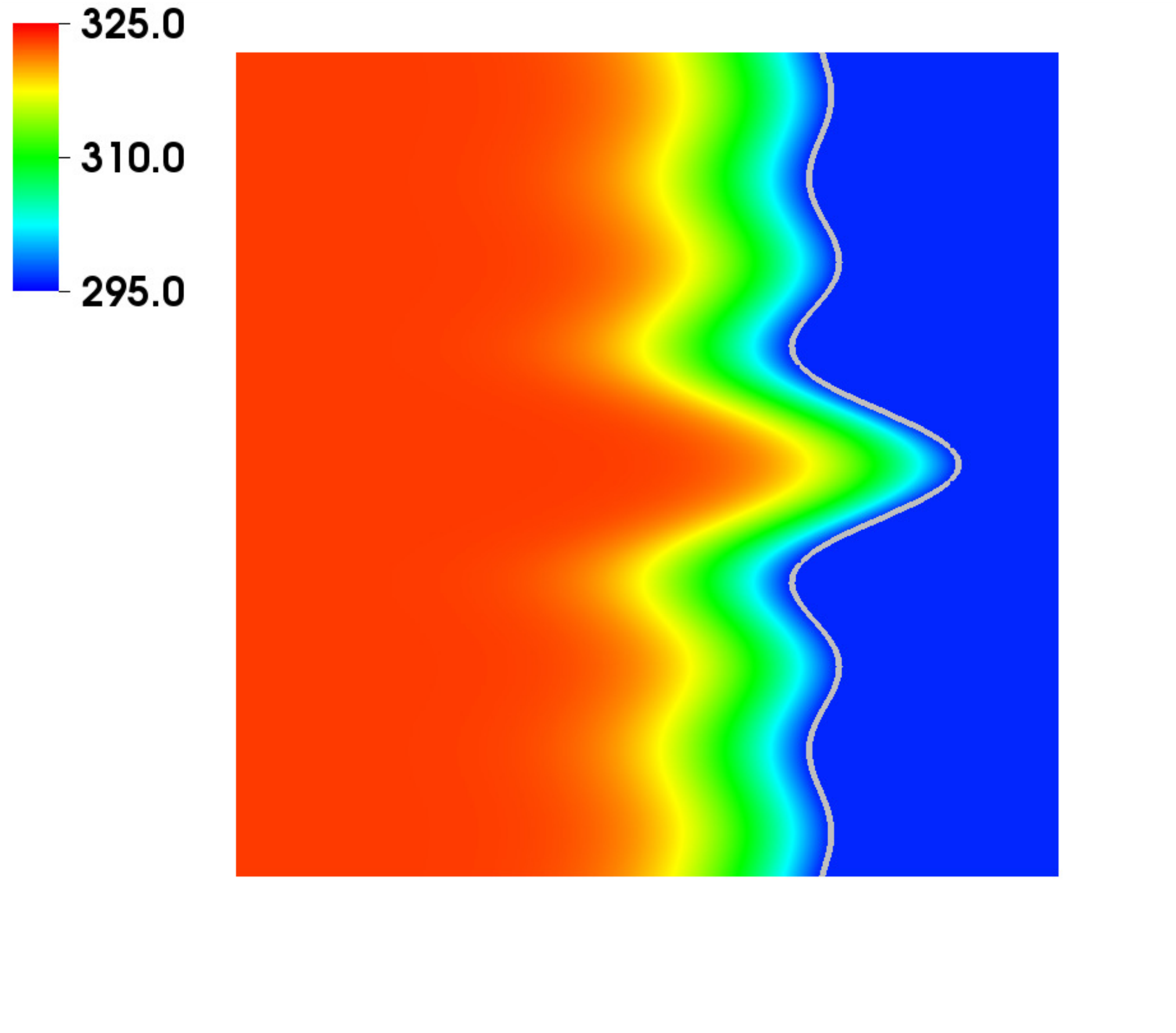}
    \caption{Pseudocolor of gas temperature [K].}\label{fig:2Dinterface:gY}
  \end{subfigure}
    \begin{subfigure}{0.49\textwidth}
    \centering
    \includegraphics[width=\linewidth]{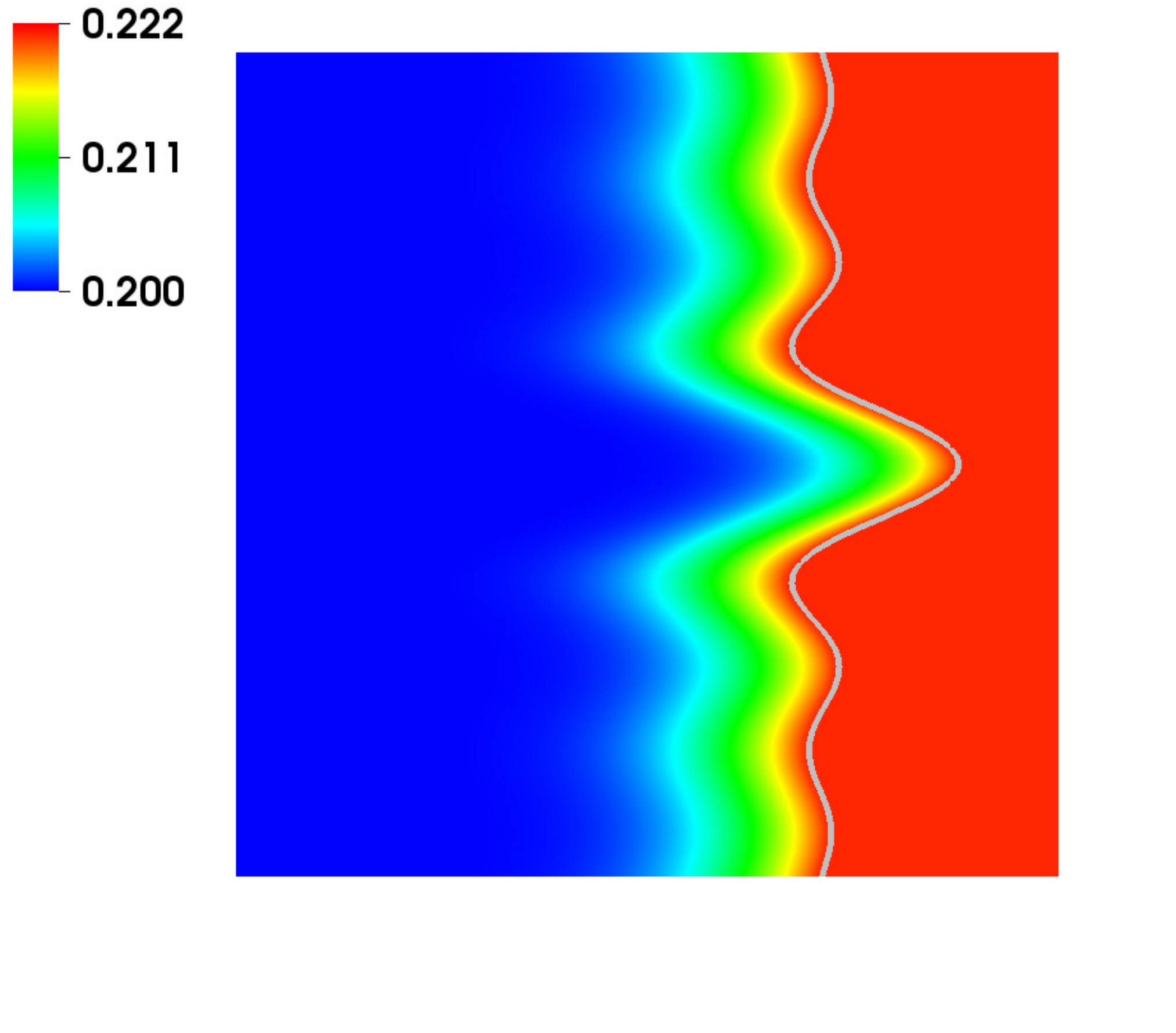}
    \caption{Pseudocolor of vapor mass fraction.}\label{fig:2Dinterface:gY}
    \end{subfigure}
    \caption{Initial field used in 2D curved interface case here shown on the \(200^2\) mesh. The interface is shown as a gray line.}\label{fig:2Dinterface}
\end{figure}

The interface shape and initial conditions are depicted in \cref{fig:2Dinterface}. The interface shape is parameterized as a function of the \(y\) position and is given by
\begin{align}
  x_\Gamma=L_x\left(0.04+w_h+w_a\left(\frac{2\pi w_ny}{L_y}\right)^{-1}\sin\left(\frac{2\pi w_ny}{L_y}\right)\right)\label{eq:2DG}
\end{align}
where \(w_h\), \(w_a\), and \(w_n\) are arbitrary constants here chosen as \(\frac{2}{3}\), \(\frac{1}{6}\), and \(5\), respectively. The grid length in the \(x\) and \(y\) directions are represented by \(L_x\) and \(L_y\), respectively and are equal \(L_x=L_y=1e-3\). The initial conditions used in this problem are chosen arbitrarily, but have been inspired by those of the previous section. The condition may be written as
\begin{align}
  T_G&=T_\Gamma +(T_\infty-T_\Gamma )\erf\left(5\frac{x_\Gamma-x}{L_x}\right) \\
  Y_G&=Y_\Gamma +(Y_\infty-Y_\Gamma )\erf\left(5\frac{x_\Gamma-x}{L_x}\right) \\
  T_L&=T_\Gamma.
\end{align}
The fluid properties are the same as the previous cases and are listed in \cref{tab:Neumann}.

\subsection{Solution Convergence} 
\begin{figure}
  \centering
  \includegraphics[height=\linewidth,angle=90]{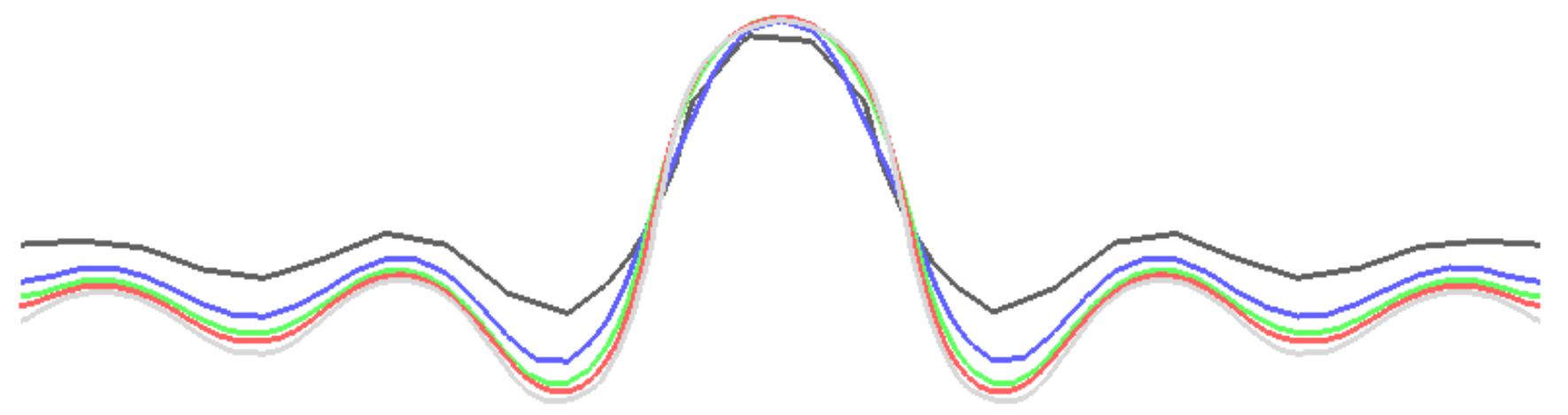}
  \caption{Interface convergence using REA2. The grids \(25\times 25\), \(50\times 50\), \(75\times 75\), \(100\times 100\), and \(200\times 200\) are represented using black, blue, green, red, and gray, respectively.}\label{fig:2DconvergeInt}
\end{figure}

\begin{figure}
  %% \centering
  \begin{subfigure}{0.49\textwidth}
    \centering
    \includetikz{\linewidth}{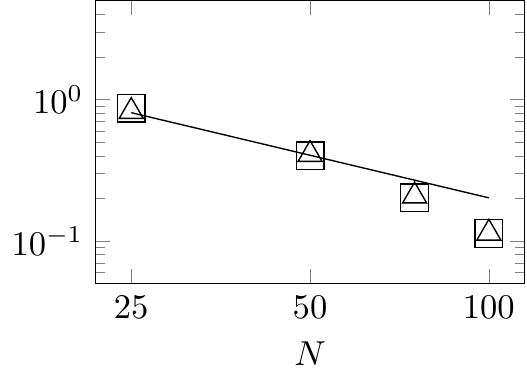}
    \caption{\(L_1\) error in gas temperature.}\label{fig:2D:L1}
  \end{subfigure}
  \begin{subfigure}{0.49\textwidth}
    \centering
    \includetikz{\linewidth}{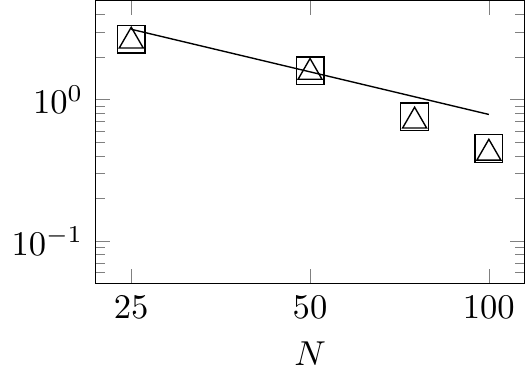}
    \caption{\(L_\infty\) error in gas temperature.}\label{fig:2D:Linf}
  \end{subfigure}
  \begin{subfigure}{0.49\textwidth}
    \centering
    \includetikz{\linewidth}{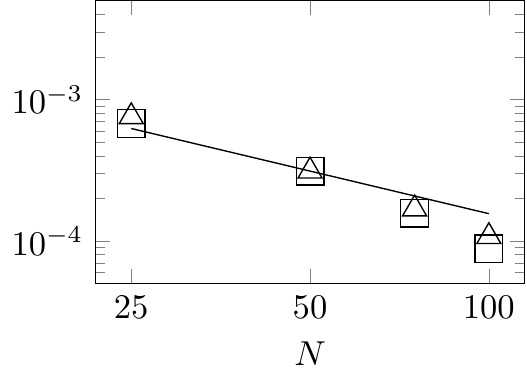}
    \caption{\(L_1\) error in vapor mass fraction.}\label{fig:2D:L1Vap}
  \end{subfigure}
  \begin{subfigure}{0.49\textwidth}
    \centering
    \includetikz{\linewidth}{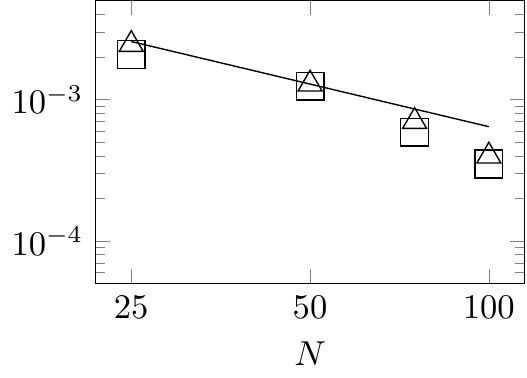}
    \caption{\(L_\infty\) error in vapor mass fraction.}\label{fig:2D:LinfVap}
  \end{subfigure}
  \begin{subfigure}{0.49\textwidth}
    \centering
    \includetikz{\linewidth}{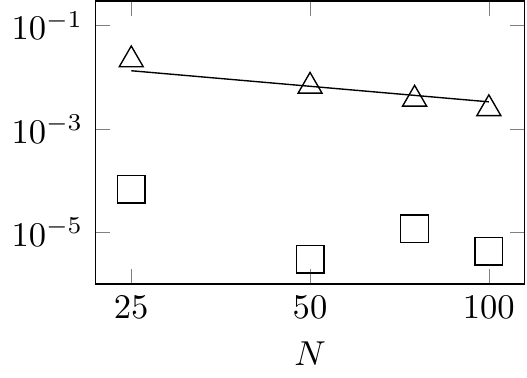}
    \caption{\(L_1\) error in liquid temperature.}\label{fig:2D:L1Liq}
  \end{subfigure}
  \begin{subfigure}{0.49\textwidth}
    \centering
    \includetikz{\linewidth}{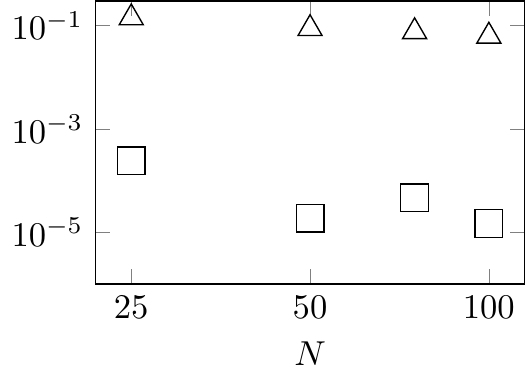}
    \caption{\(L_\infty\) error in liquid temperature.}\label{fig:2D:LinfLiq}
  \end{subfigure}
  \begin{subfigure}{0.49\textwidth}
    \centering
    \includetikz{\linewidth}{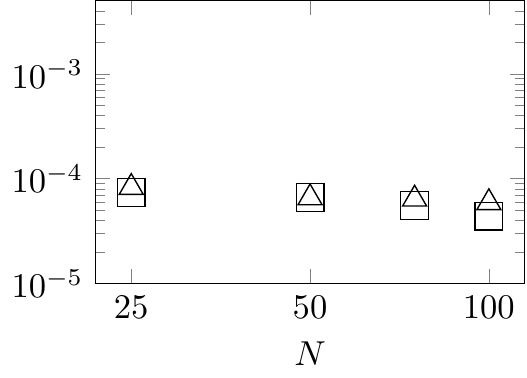}
    \caption{\(L_1\) error in liquid velocity.}\label{fig:2D:LinfVelo}
  \end{subfigure}
  \begin{subfigure}{0.49\textwidth}
    \centering
    \includetikz{\linewidth}{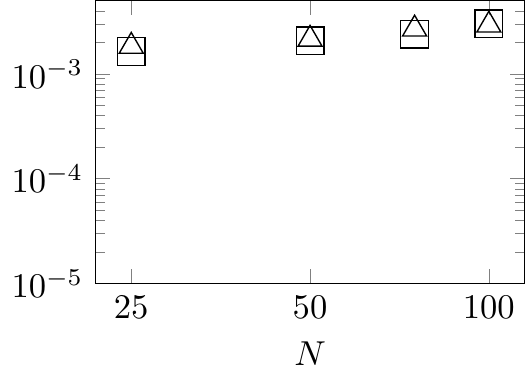}
    \caption{\(L_\infty\) error in liquid velocity.}\label{fig:2D:LinfVelo}
  \end{subfigure}
  \caption{Case 3. Errors computed at \(t=0.1\text{ s}\). MB (\reftikz{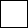}{tikz:2DVelo_Linf:MaBo}), REA2 (\reftikz{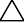}{tikz:2DVelo_Linf:REAL2}). A first order trend line has been added where appropriate.}\label{fig:2D}
\end{figure}

Simulations are performed for grids of size \(25\times 25\), \(50\times 50\), \(75\times 75\), and \(100\times 100\), and the data are compared to a simulation with grid size \(200\times 200\) for accuracy. Simulation data are taken at time \(t=0.05s\). Both the MB and REA2 techniques have been used for this study, and both technique perform well for grid convergence. Images of the convergence of the interface position are shown in \cref{fig:2DconvergeInt}. The profile seems to sharpen under grid refinement. The center most hump appears taller, whereas the outer humps appear to decrease in height consistent with the principle of mass conservation. The interface converges rapidly towards the fine mesh profile. \Cref{fig:2D} depicts the convergence of the gas temperature, vapor mass fraction, liquid temperature, and liquid velocity in the domain. A first order scaling is recovered for both \(L_1\) and \(L_\infty\) errors in the gas fields. For the liquid temperature field, first order convergence is seen in \(L_1\), but slower convergence is noted for \(L_\infty\). The liquid velocity field appears to converge at a very slow rate in \(L_1\), but diverges in \(L_\infty\). This is not entirely surprising. From the previous section, we noted that the liquid velocity is directly proportional to \(\dot m\) and therefore directly proportional to the scalar gradient at the interface. Since the scalar field is itself only first order accurate, its gradient would be expected to lose an order of accuracy, recovering only zeroth order accuracy. This explains the lack of convergence in the liquid velocity field. Since the velocity controls the liquid motion and the rate of heat advection into and out of the interface, it is reasonable that this would have an impact on the accuracy in the liquid temperature field as well. However, since the gas is stationary, only the boundary effect is felt by the gas, decreasing the impact of the liquid velocity on its accuracy.

It is important to note the significance of the trends in \cref{fig:2D}. First, it can be seen that the effective accuracy of the simulation in two dimensions is somewhat lower than what was recovered in one dimension. This is unsurprising because of the greater geometric complexity of the interface in two dimensions. The resolution of this complex interface shape becomes a limiting factor in the resolution of the fluid flow. Second, it can be seen that the error effect is particularly strong when measured in the \(L_\infty\) sense. It is common within the literature to report errors in terms of smooth error norms such as \(L_1\). This form of measurement can be misleading when numerical errors are not distributed uniformly, and that is the case for these simulations where errors are largest near the interface. This calls into question the use of smooth error norms for numerical analysis of flows with sharp features or discontinuities.

While the current work fails to convergence in \(L_\infty\), it does converge in the more traditional \(L_1\) sense. Consistent with the trends in the literature, the remainder of this paper will assume that the recovered \(L_1\) convergence is sufficient to perform exploratory studies of vaporizing multiphase flows. However, future work will focus on improving the accuracy of the flow solver, so that improved convergence will can be recovered.

\subsection{Temporal Stability and Convergence}\label{sec:stability}
The temporal convergence and stability of the solver is tested by performing simulations on the \(100\times 100\) grid using various time steps. The time steps are chosen to yield \(C_D=C_\lambda \in \{10,25,50,100,200\}\). The two time integration schemes are compared using both the MB and REA2 techniques. The data are summarized in \cref{tab:2Dsemi,tab:2Dfull}.

Using the semi-implicit time integrator, similar trends are seen in errors for gas temperature and vapor mass fraction using both MB and REA2 techniques. Both \(L_1\) and \(L_\infty\) errors decrease with \(C_\lambda\). There is a noticeable difference in the error of the liquid temperature and velocity, as has been seen in the previous sections. It is interesting to note that neither of these field appears to converge under temporal refinement (decreasing \(C_\lambda\)) according to the \(L_1\) metric. Surprisingly, the liquid velocity converges in both cases under the \(L_\infty\) metric. It should also be noted that for the REA2 technique, there is spike in solution error at \(C_\lambda=200\). It will later be demonstrated that this is due to an issue with solution monotonicity.

Using the fully implicit time integrator a few trends can be noted.  The MB technique seems to recover superior temporal convergence using the fully implicit solver than the semi-implicit solver. Ignoring the \(C_\lambda=200\) case of REA2 technique, their errors are largely the same using both time integrators. This is not a surprise. While the semi-implicit solver has higher formal accuracy than the fully implicitly integrator, this property is no use when the leading error term in the flow is dominated by the first order treatment of the interfacial diffusion flux~\cref{sec:flux}.

\begin{table}
  \begin{subfigure}{0.49\textwidth}
    \centering
    \begin{tabular}{r c c}
      \toprule
      \(C_\lambda\) & MB & REA2 \\
      \midrule
      \(10\)  & \(1.36e-1\) & \(1.36e-1\) \\ 
      \(25\)  & \(1.65e-1\) & \(1.66e-1\) \\ 
      \(50\)  & \(1.89e-1\) & \(1.95e-1\) \\ 
      \(100\) & \(2.06e-1\) & \(2.21e-1\) \\
      \(200\) & \(2.22e-1\) & \(7.25e-1\) \\
      \bottomrule
    \end{tabular}
    \caption{\(L_1\) error in gas temperature.}
  \end{subfigure}
  \begin{subfigure}{0.49\textwidth}
    \centering
    \begin{tabular}{r c c}
      \toprule
      \(C_\lambda\) & MB & REA2 \\
      \midrule
      \(10\)  & \(5.66e-1\) & \(5.11e-1\) \\ 
      \(25\)  & \(7.48e-1\) & \(6.97e-1\) \\ 
      \(50\)  & \(9.48e-1\) & \(9.17e-1\) \\ 
      \(100\) & \(1.09e-0\) & \(1.11e-0\) \\
      \(200\) & \(1.19e-0\) & \(2.42e+1\) \\
      \bottomrule   
    \end{tabular}
    \caption{\(L_\infty\) error in gas temperature.}
  \end{subfigure}
  \begin{subfigure}{0.49\textwidth}
    \centering
    \begin{tabular}{r c c}
      \toprule
      \(C_\lambda\) & MB & REA2 \\
      \midrule
      \(10\)  & \(1.07e-4\) & \(1.21e-4\) \\ 
      \(25\)  & \(1.32e-4\) & \(1.39e-4\) \\ 
      \(50\)  & \(1.49e-4\) & \(1.62e-4\) \\ 
      \(100\) & \(1.55e-4\) & \(1.82e-4\) \\
      \(200\) & \(1.60e-4\) & \(1.67e-3\) \\
      \bottomrule
    \end{tabular}
    \caption{\(L_1\) error in vapor mass fraction.}
  \end{subfigure}
  \begin{subfigure}{0.49\textwidth}
    \centering
    \begin{tabular}{r c c}
      \toprule
      \(C_\lambda\) & MB & REA2 \\
      \midrule
      \(10\)  & \(4.43e-4\) & \(4.91e-4\) \\ 
      \(25\)  & \(5.92e-4\) & \(6.32e-4\) \\ 
      \(50\)  & \(7.46e-4\) & \(7.68e-4\) \\ 
      \(100\) & \(8.35e-4\) & \(8.94e-4\) \\
      \(200\) & \(8.84e-4\) & \(1.54e-2\) \\
      \bottomrule   
    \end{tabular}
    \caption{\(L_\infty\) error in vapor mass fraction.}
  \end{subfigure}
  \begin{subfigure}{0.49\textwidth}
    \centering
    \begin{tabular}{r c c}
      \toprule
      \(C_\lambda\) & MB & REA2 \\
      \midrule
      \(10\)  & \(1.59e-4\) & \(2.39e-3\) \\ 
      \(25\)  & \(4.49e-4\) & \(2.33e-3\) \\ 
      \(50\)  & \(2.17e-4\) & \(2.59e-3\) \\ 
      \(100\) & \(8.63e-4\) & \(2.92e-3\) \\
      \(200\) & \(1.47e-3\) & \(2.39e-1\) \\
      \bottomrule
    \end{tabular}
    \caption{\(L_1\) error in liquid temperature.}
  \end{subfigure}
  \begin{subfigure}{0.49\textwidth}
    \centering
    \begin{tabular}{r c c}
      \toprule
      \(C_\lambda\) & MB & REA2 \\
      \midrule
      \(10\)  & \(5.73e-4\) & \(5.61e-2\) \\ 
      \(25\)  & \(1.82e-3\) & \(5.78e-2\) \\ 
      \(50\)  & \(1.48e-3\) & \(6.39e-2\) \\ 
      \(100\) & \(6.03e-3\) & \(5.77e-2\) \\
      \(200\) & \(1.54e-2\) & \(2.37e-0\) \\
      \bottomrule   
    \end{tabular}
    \caption{\(L_\infty\) error in liquid temperature.}
  \end{subfigure}
    \begin{subfigure}{0.49\textwidth}
    \centering
    \begin{tabular}{r c c}
      \toprule
      \(C_\lambda\) & MB & REA2 \\
      \midrule
      \(10\)  & \(3.59e-5\) & \(3.12e-5\) \\ 
      \(25\)  & \(4.86e-5\) & \(4.17e-5\) \\ 
      \(50\)  & \(4.57e-5\) & \(4.01e-5\) \\ 
      \(100\) & \(4.51e-5\) & \(3.79e-5\) \\
      \(200\) & \(4.28e-5\) & \(2.11e-4\) \\
      \bottomrule
    \end{tabular}
    \caption{\(L_1\) error in liquid velocity.}
  \end{subfigure}
  \begin{subfigure}{0.49\textwidth}
    \centering
    \begin{tabular}{r c c}
      \toprule
      \(C_\lambda\) & MB & REA2 \\
      \midrule
      \(10\)  & \(1.70e-4\) & \(2.24e-4\) \\ 
      \(25\)  & \(1.50e-4\) & \(2.25e-4\) \\ 
      \(50\)  & \(1.86e-4\) & \(2.77e-4\) \\ 
      \(100\) & \(1.91e-4\) & \(2.89e-4\) \\
      \(200\) & \(1.94e-4\) & \(5.06e-4\) \\
      \bottomrule   
    \end{tabular}
    \caption{\(L_\infty\) error in liquid velocity.}
  \end{subfigure}
  \caption{Errors using the semi-implicit time integrator.}\label{tab:2Dsemi}
\end{table}

\begin{table}
  \begin{subfigure}{0.49\textwidth}
    \centering
    \begin{tabular}{r c c}
      \toprule
      \(C_\lambda\) & MB & REA2 \\
      \midrule
      \(10\)  & \(1.36e-1\) & \(1.36e-1\) \\ 
      \(25\)  & \(1.66e-1\) & \(1.66e-1\) \\ 
      \(50\)  & \(1.91e-1\) & \(1.94e-1\) \\ 
      \(100\) & \(2.12e-1\) & \(2.19e-1\) \\
      \(200\) & \(2.27e-1\) & \(2.39e-1\) \\
      \bottomrule
    \end{tabular}
    \caption{\(L_1\) error in gas temperature.}
  \end{subfigure}
  \begin{subfigure}{0.49\textwidth}
    \centering
    \begin{tabular}{r c c}
      \toprule
      \(C_\lambda\) & MB & REA2 \\
      \midrule
      \(10\)  & \(5.66e-1\) & \(5.00e-1\) \\ 
      \(25\)  & \(7.48e-1\) & \(6.94e-1\) \\ 
      \(50\)  & \(9.53e-1\) & \(9.13e-1\) \\ 
      \(100\) & \(1.11e-0\) & \(1.10e-0\) \\
      \(200\) & \(1.20e-0\) & \(1.18e-0\) \\
      \bottomrule   
    \end{tabular}
    \caption{\(L_\infty\) error in gas temperature.}
  \end{subfigure}
  \begin{subfigure}{0.49\textwidth}
    \centering
    \begin{tabular}{r c c}
      \toprule
      \(C_\lambda\) & MB & REA2 \\
      \midrule
      \(10\)  & \(1.06e-4\) & \(1.21e-4\) \\ 
      \(25\)  & \(1.31e-4\) & \(1.37e-4\) \\ 
      \(50\)  & \(1.51e-4\) & \(1.58e-4\) \\ 
      \(100\) & \(1.70e-4\) & \(1.76e-4\) \\
      \(200\) & \(1.83e-4\) & \(1.94e-4\) \\
      \bottomrule
    \end{tabular}
    \caption{\(L_1\) error in vapor mass fraction.}
  \end{subfigure}
  \begin{subfigure}{0.49\textwidth}
    \centering
    \begin{tabular}{r c c}
      \toprule
      \(C_\lambda\) & MB & REA2 \\
      \midrule
      \(10\)  & \(4.42e-4\) & \(4.90e-4\) \\ 
      \(25\)  & \(5.87e-4\) & \(6.31e-4\) \\ 
      \(50\)  & \(7.51e-4\) & \(7.59e-4\) \\ 
      \(100\) & \(8.79e-4\) & \(8.76e-4\) \\
      \(200\) & \(9.54e-4\) & \(9.60e-4\) \\
      \bottomrule   
    \end{tabular}
    \caption{\(L_\infty\) error in vapor mass fraction.}
  \end{subfigure}
  \begin{subfigure}{0.49\textwidth}
    \centering
    \begin{tabular}{r c c}
      \toprule
      \(C_\lambda\) & MB & REA2 \\
      \midrule
      \(10\)  & \(7.07e-5\) & \(2.37e-3\) \\ 
      \(25\)  & \(2.39e-4\) & \(2.28e-3\) \\ 
      \(50\)  & \(4.76e-4\) & \(2.39e-3\) \\ 
      \(100\) & \(7.47e-4\) & \(2.49e-3\) \\
      \(200\) & \(9.79e-4\) & \(2.85e-3\) \\
      \bottomrule
    \end{tabular}
    \caption{\(L_1\) error in liquid temperature.}
  \end{subfigure}
  \begin{subfigure}{0.49\textwidth}
    \centering
    \begin{tabular}{r c c}
      \toprule
      \(C_\lambda\) & MB & REA2 \\
      \midrule
      \(10\)  & \(2.46e-4\) & \(5.59e-2\) \\ 
      \(25\)  & \(8.52e-4\) & \(5.76e-2\) \\ 
      \(50\)  & \(1.75e-3\) & \(6.26e-2\) \\ 
      \(100\) & \(2.93e-3\) & \(5.61e-2\) \\
      \(200\) & \(4.18e-3\) & \(5.93e-2\) \\
      \bottomrule   
    \end{tabular}
    \caption{\(L_\infty\) error in liquid temperature.}
  \end{subfigure}
    \begin{subfigure}{0.49\textwidth}
    \centering
    \begin{tabular}{r c c}
      \toprule
      \(C_\lambda\) & MB & REA2 \\
      \midrule
      \(10\)  & \(3.59e-5\) & \(2.69e-5\) \\ 
      \(25\)  & \(4.87e-5\) & \(4.20e-5\) \\ 
      \(50\)  & \(4.56e-5\) & \(4.04e-5\) \\ 
      \(100\) & \(4.55e-5\) & \(3.83e-5\) \\
      \(200\) & \(4.51e-5\) & \(5.23e-5\) \\
      \bottomrule
    \end{tabular}
    \caption{\(L_1\) error in liquid velocity.}
  \end{subfigure}
  \begin{subfigure}{0.49\textwidth}
    \centering
    \begin{tabular}{r c c}
      \toprule
      \(C_\lambda\) & MB & REA2 \\
      \midrule
      \(10\)  & \(1.69e-4\) & \(2.59e-4\) \\ 
      \(25\)  & \(1.49e-4\) & \(2.25e-4\) \\ 
      \(50\)  & \(1.75e-4\) & \(2.64e-4\) \\ 
      \(100\) & \(1.85e-4\) & \(2.84e-4\) \\
      \(200\) & \(1.94e-4\) & \(2.41e-4\) \\
      \bottomrule   
    \end{tabular}
    \caption{\(L_\infty\) error in liquid velocity.}
  \end{subfigure}
  \caption{Errors using the fully implicit time integrator.}\label{tab:2Dfull}
\end{table}

\subsubsection{Monotonicity and Stability}
\begin{figure}
  \centering
  \begin{subfigure}{0.49\textwidth}
    \includegraphics[width=\linewidth]{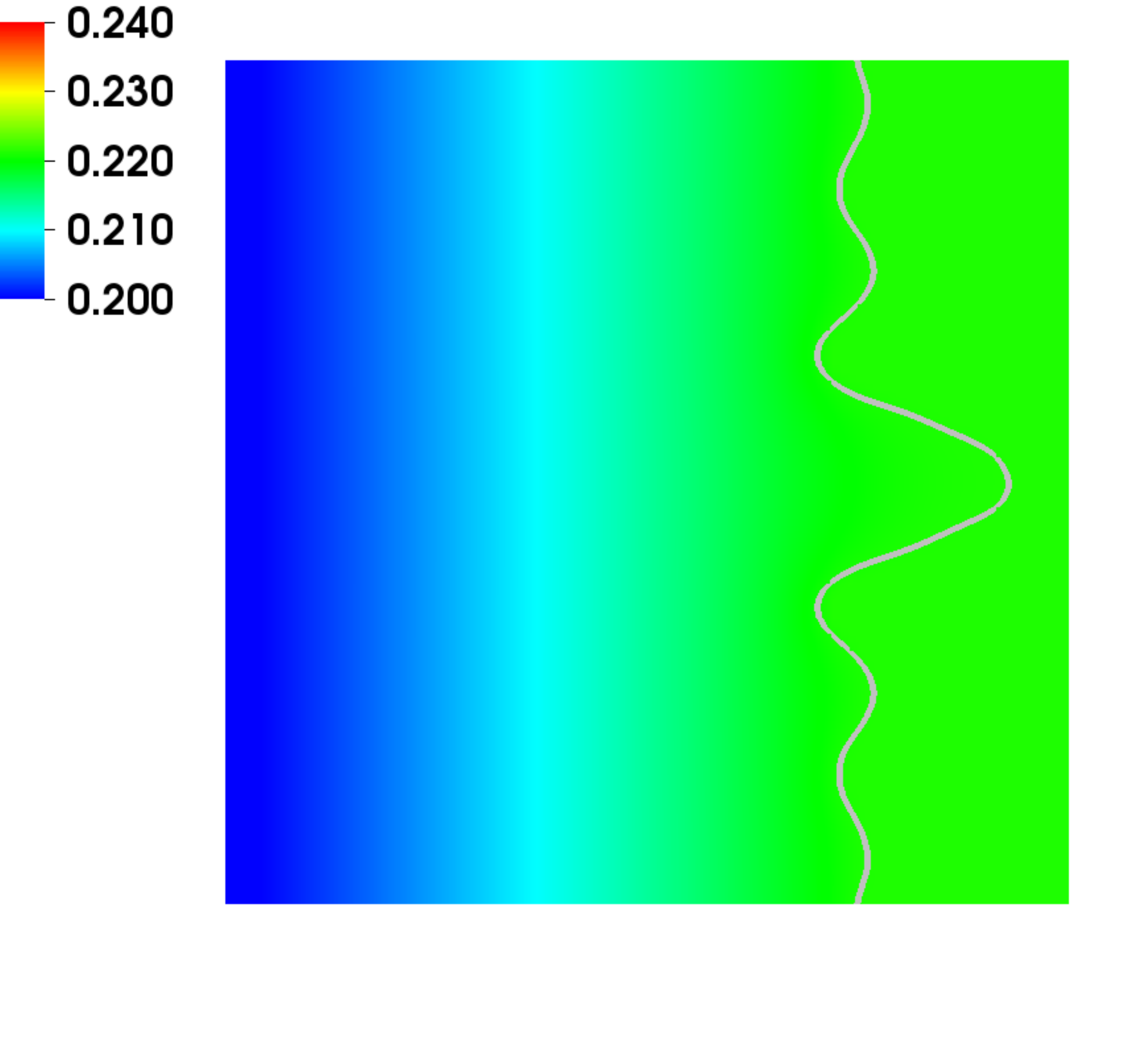}
    \caption{Pseudocolor of vapor mass fraction. Fully implicit}
  \end{subfigure}
  \begin{subfigure}{0.49\textwidth}
    \includegraphics[width=\linewidth]{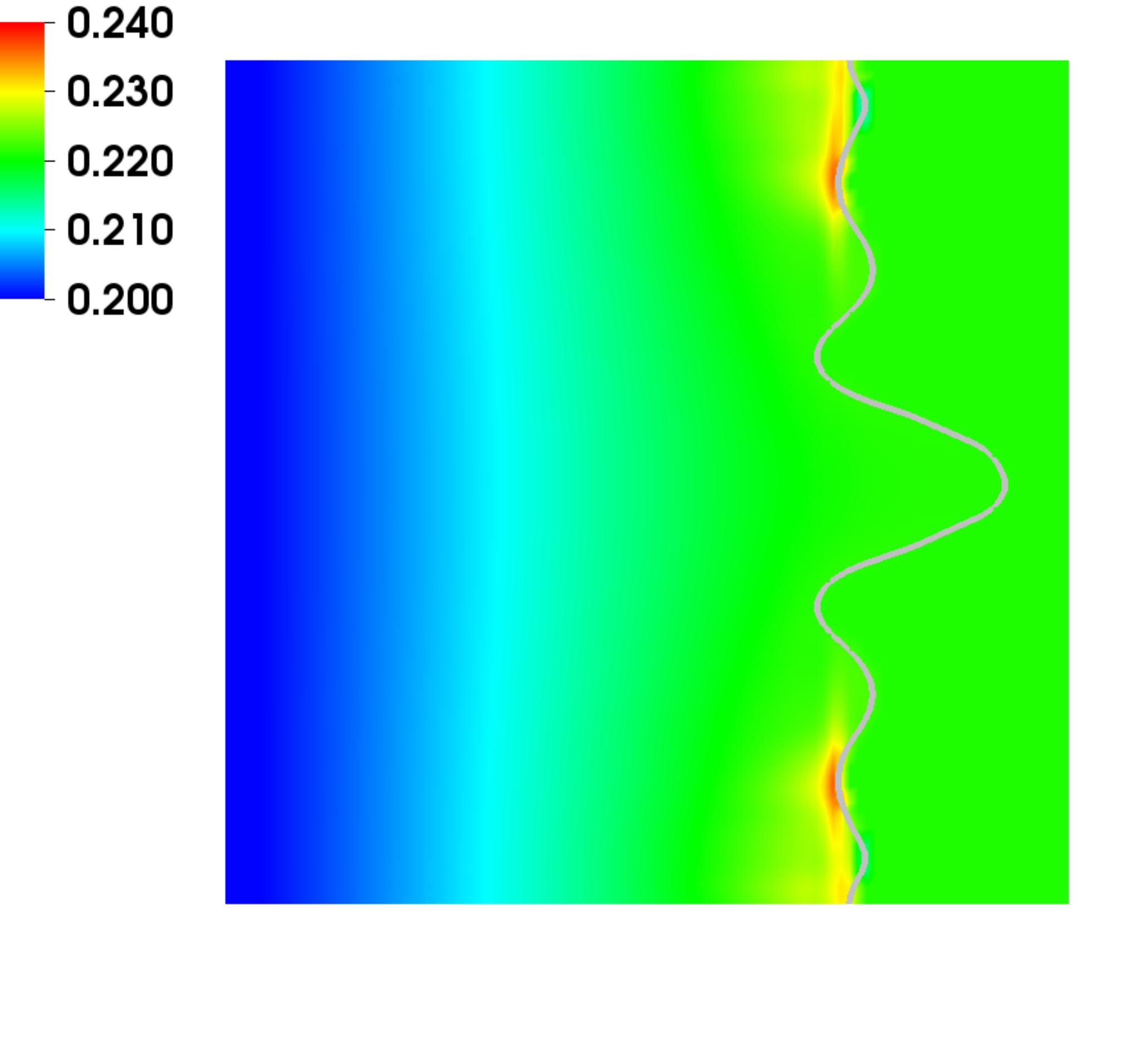}
    \caption{Pseudocolor of vapor mass fraction. Semi implicit}
  \end{subfigure}
  \begin{subfigure}{0.49\textwidth}
    \includegraphics[width=\linewidth]{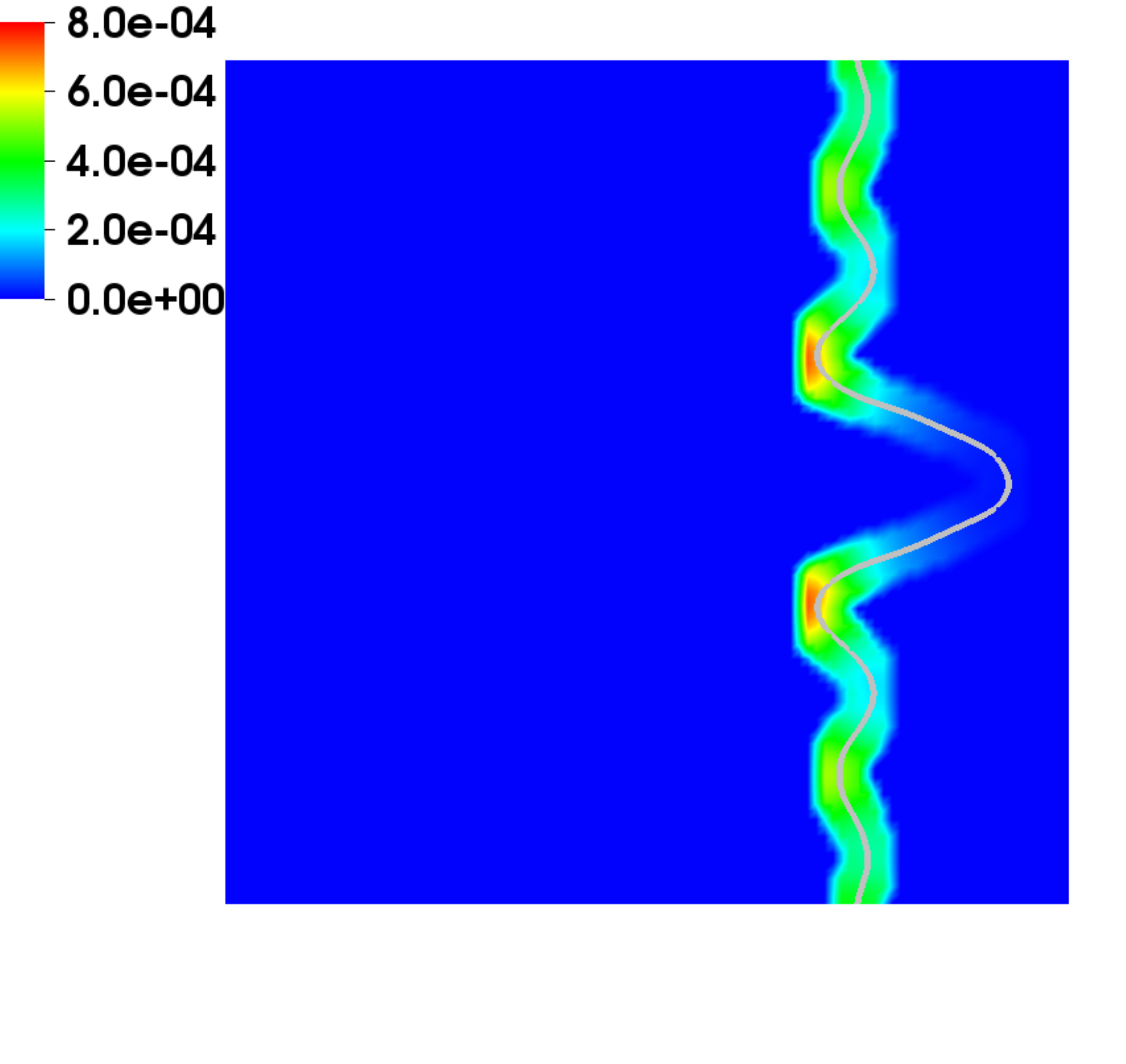}
    \caption{Pseudocolor of \(\dot m\). Fully implicit}
  \end{subfigure}
  \begin{subfigure}{0.49\textwidth}
    \includegraphics[width=\linewidth]{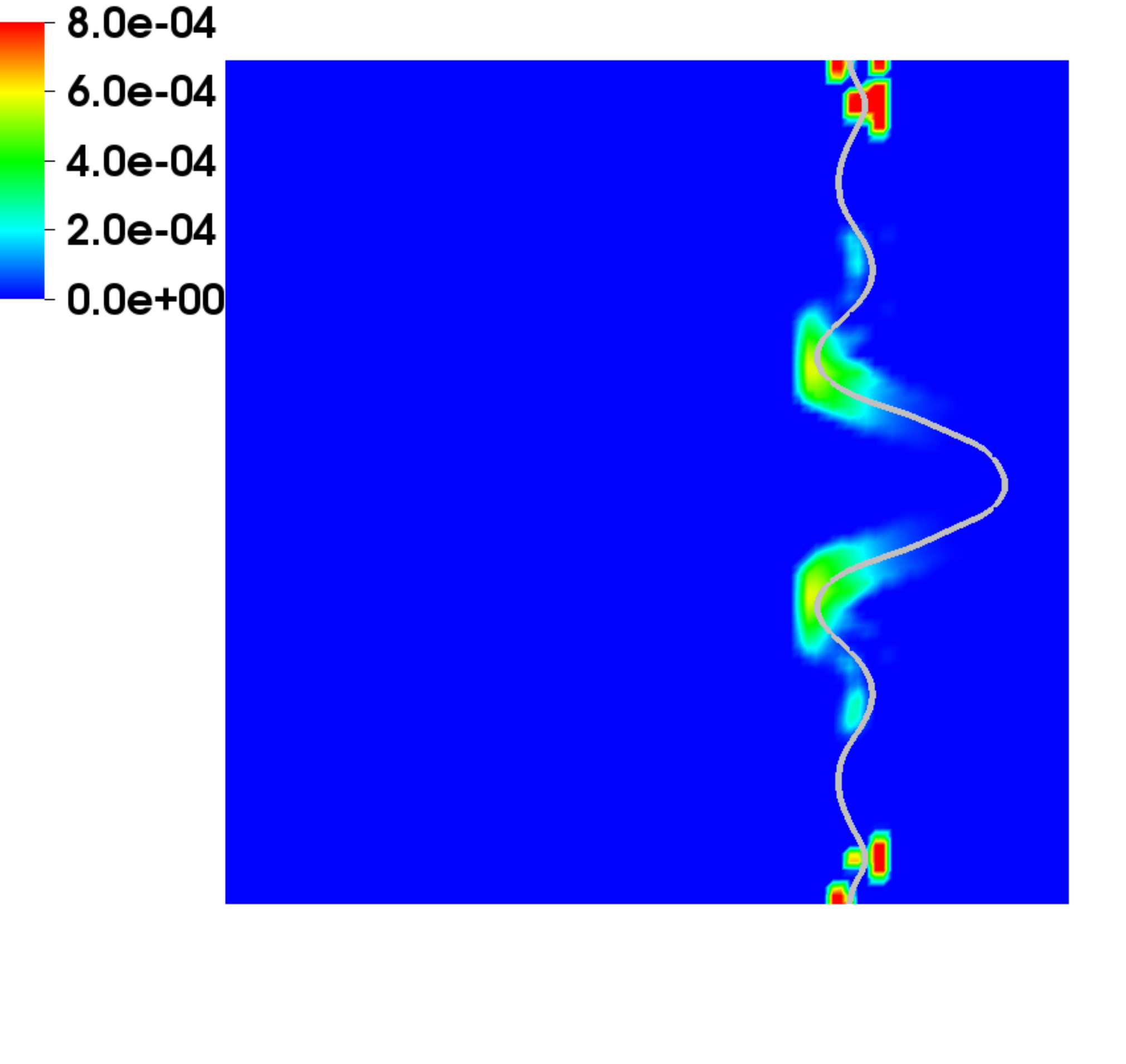}
    \caption{Pseudocolor of \(\dot m\). Semi implicit}
  \end{subfigure}
  \caption{Influence of time integration on field data using fully implicit and semi implicit solver. Data taken at simulation end using \(C=200\). The interface is shown as a gray line.}\label{fig:wiggle}
\end{figure}

\begin{figure}
  \centering
  \begin{subfigure}{0.49\textwidth}
    \includetikz{\linewidth}{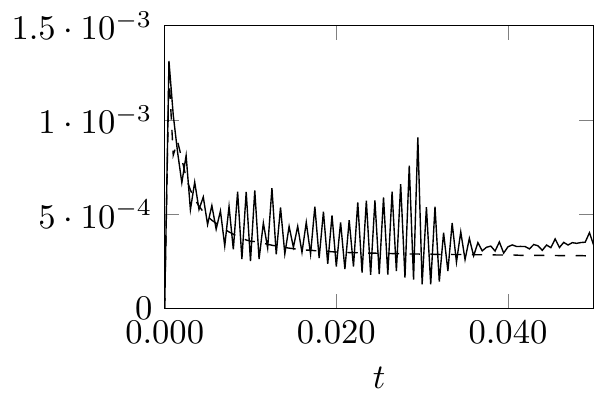}
    \caption{\(\dot m\).}\label{fig:mono_mdot}
  \end{subfigure}
  \begin{subfigure}{0.49\textwidth}
    \includetikz{\linewidth}{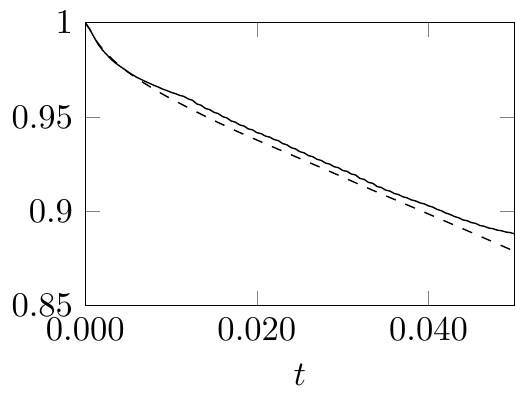}
    \caption{Normalized volume, \(V/V_0\).}\label{fig:mono_vol}
  \end{subfigure}
  \caption{Influence of time integration on temporal trace of data using fully implicit and semi implicit solver.  Semi implicit (\reftikz{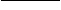}{tikz:time_volume:tridiag}), Fully implicit (\reftikz{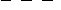}{tikz:time_volume:full}).}
\end{figure}

The data suggested both MB and REA2 schemes recovered better stability and convergence properties using the fully implicit time integration scheme. This follows from the fact that this scheme is a monotone scheme, and therefore disallows the development of numerical oscillation developing in the flow. \Cref{fig:wiggle} demonstrates snapshots of the vapor mass fraction and \(\dot m\) fields at simulation end using the REA2 technique and \(C_\lambda=200\). Using the semi-implicit time integration a region of high vapor mass fraction appears near the interface leading to a non-monotonic variation of the field. This is not seen in the fully implicit time integrator. As a result, the \(\dot m\) fields look very different. The fully implicit scheme demonstrates smooth variation of \(\dot m\) , whereas the semi-implicit scheme demonstrates chunking. The effect can also be seen by monitoring the 
temporal behavior of \(\dot m\) in \cref{fig:mono_mdot}. While \(\dot m\) varies smoothly for the fully implicit scheme, it is oscillatory for the semi-implicit scheme. This translates to rapid changes in the slope of the normalized volume (\cref{fig:mono_vol}), and the eventual divergence of the solution.

\section{Vaporization of an Isolated Spherical Droplet}\label{sec:3D}
%% \subsection{The \(d^2\) Law}\label{sec:d2}
The previous  solutions, while interesting for
verification, have little practical value. A somewhat more physically relevant problem
is the problem of vaporizing a spherical droplet in a quiescent flow. Assuming
the problem is quasi-steady (temporal derivatives of the scalars and momentum
are negligible) and the gas obeys the continuity equation, \(\nabla\cdot\bm u=0\), the solution to the temperature and mass fraction fields can be found analytically, and is given in any text on the subject \cite{Sirignano2010}. The remarkable aspect of the solution is that the analytical result predicts that the squared diameter of the droplet (\(d^2\)) varies linearly in time. This result is also observed experimentally for both vaporization and combustion \cite{Law1982}.

We follow the derivation of Rueda-Villegas et al. \cite{Villegas2016} who obtain the solution
\begin{align}
  \frac{T-T_\Gamma+L/C}{T_\infty-T_\Gamma+L/C}=&e^{-\frac{QC_p}{2\pi k}\frac{1}{d}}, \text{ and} \label{eq:D2_T}\\
  \frac{Y-1}{Y_\infty-1}=&e^{-\frac{Q}{2\pi\rho D}\frac{1}{d}}. \label{eq:D2_Y}
\end{align}
In \cref{eq:D2_T,eq:D2_Y}, \(Q\) is a constant related to \(\dot m\). The values \(\dot m\), \(Y_\Gamma\), and \(T_\Gamma\) are related through
\begin{align}
  %% Q = \frac{2\pi dk}{C_p}\ln{\frac{T_\infty-T_\Gamma+L/C}{L}} \label{eq:D2_mdot}\\
  Q = -2\pi d\rho_GD\ln\left(\frac{Y-1}{Y_\infty -1}\right), \text{ and} \label{eq:D2_mdot}\\
  T_\Gamma=T_\infty+\frac{L}{C_p}\left(1-{\left(\frac{Y_\infty-1}{Y_\Gamma-1}\right)}^{D/\lambda}\right),\label{eq:D2_link}
\end{align}
which is derived form \cref{eq:D2_T,eq:D2_Y}. \Cref{eq:D2_mdot} is more commonly written in the form
\begin{align}
  Q = 2\pi d\rho_GD\ln\left(1+B_M\right)
\end{align}
where \(B_M=\frac{Y_\Gamma-Y_\infty}{1-Y_\Gamma}\) is the mass transfer number. This relation can also be used to compute the vaporization rate per unit area, yielding
\begin{align}
  \dot m \equiv \frac{Q}{\pi d^2}= \frac{2}{d}\rho_GD\ln\left(1+B_M\right)
\end{align}
Note the explicit dependence of the solution on the current droplet diameter, \(d\). \Cref{eq:D2_mdot,eq:D2_link} must be solved simultaneously with the Clausius-Clapeyron relation, \cref{eq:CC}, to complete the system.

In this section, parameters are chosen to imitate the properties of acetone. It was pointed out by Raessi et al. \cite{Raessi2009}, that for practical simulation of liquid vaporization using real physical properties, the condition \(C_\sigma <1\) often becomes the limiting CFL condition. Since surface tension is not important to the mechanics of the \(d^2\) law, the surface tension was reduced to zero, allowing for the use of a larger numerical time step. Similarly, the use of the monotone, fully implicit scalar diffusion solver frees the simulation from time scales \(C_D\), \(C_{\lambda, L}\), and \(C_{\lambda ,G}\). Only the advection CFL condition need be satisfied, here forced to be \(C_u<1\) for all simulations.

Some discussion must be devoted to the implementation of the boundary conditions. The simulation is run on a periodic, cubic domain using a Cartesian solver in three dimensions. The spherical symmetry implied by the far field boundary condition is approximated by centering the domain on the droplet center of mass. Tests are performed to characterize the influence of the boundary condition on the evaporation rate (see \cref{sec:far}). It was found that at sufficiently early times in the simulation, boundary effects were negligible.

A further modification is required to deal with the boundary conditions. Experience has shown that small errors in the liquid velocity field can lead to motion of the droplet away from the center of the domain, breaking the implied symmetry of the boundary conditions. To remove this source of error from this suite of tests, \(-\bm n_\Gamma\dot m\) is used as the velocity for VOF transport. This effectively enforces the condition, \(\bm u_L=\bm 0\). However, the reconstructed \(\bm u_L\) is still used (rather than \(\bm 0\)) for the liquid temperature advection.

\begin{figure}
  \centering
  \begin{subfigure}{0.495\textwidth}
    \centering
    \includegraphics[width=\linewidth]{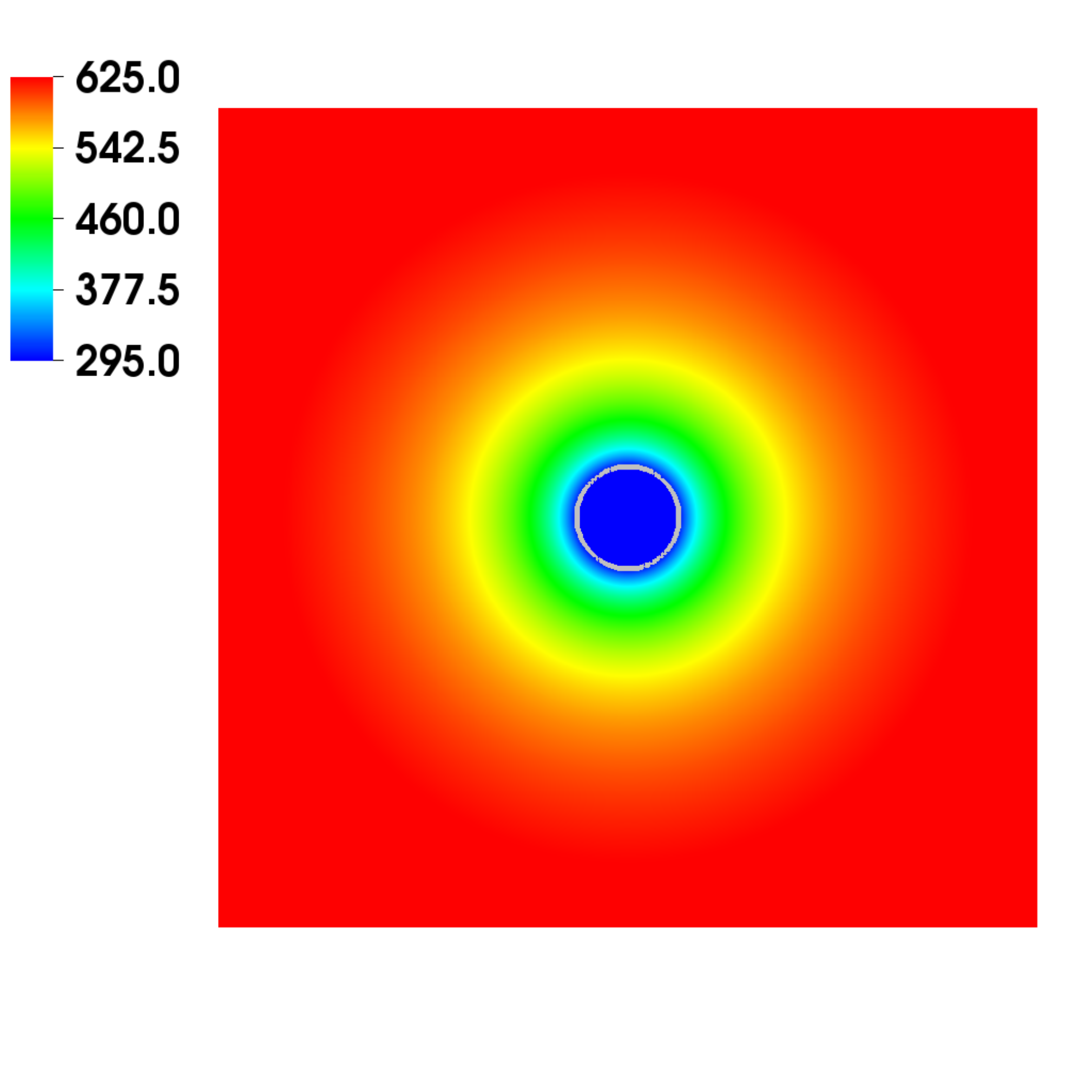}
    \caption{Pseudocolor of gas temperature [K].}\label{fig:D2_t}
  \end{subfigure}
    \begin{subfigure}{0.495\textwidth}
    \centering
    \includegraphics[width=\linewidth]{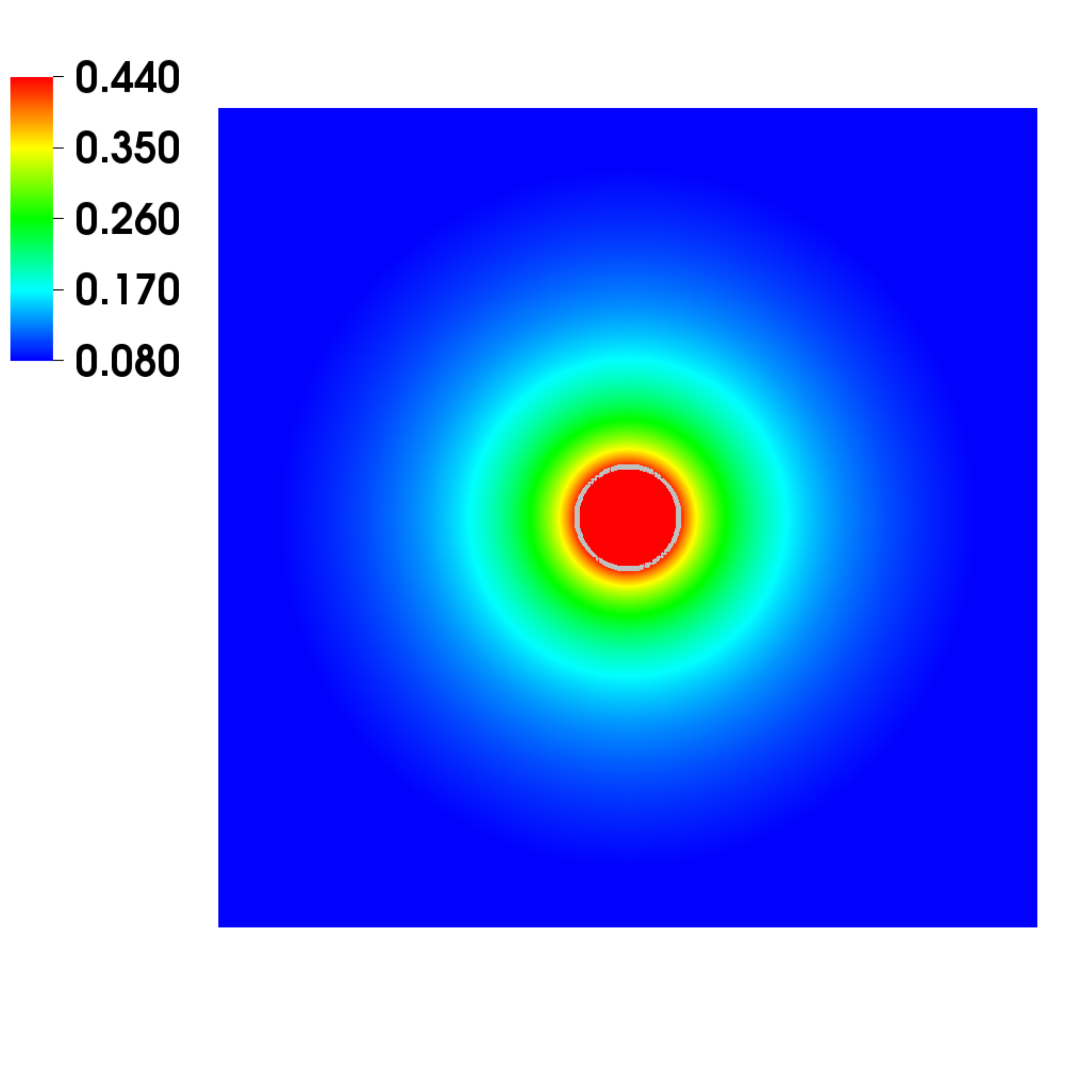}
    \caption{Pseudocolor of vapor mass fraction.}\label{fig:D2_y}
    \end{subfigure}
    \caption{Initial field used in \(d^2\) law cases, here shown on the \(128^3\) mesh. A planar cut has been taken at the center of the domain. The interface is shown as a gray line.}
\end{figure}

The chosen fluid properties are listed in \cref{tab:D2}. The actual fluid surface tension is listed, even though the value of \(\sigma=0\) is used in these simulations. The far field temperature is chosen to be \(700\text{ K}\) and the far field vapor mass fraction is \(0\). This results in interface temperature and vapor mass fraction of \(294.92\text{ K}\) and \(0.43993\), respectively. The vapor mass fraction and gas temperature fields are initialized with the analytical solution, \cref{eq:D2_T,eq:D2_Y}, and the liquid temperature field is initialized simply as \(T_\Gamma\). Note that in contrast to the simulations of previous sections, this section demonstrates a large range of values in both the temperature and vapor mass fraction fields. The initialization is demonstrated in \cref{fig:D2_t,fig:D2_y}.

A grid convergence study was performed using a domain of size \(8\times 10^{-4}\)m which corresponds to exactly eight droplet diameters. The convergence study was performed using \(N=32\), \(64\), \(128\), and \(256\) grid cells in each of the three grid directions. This corresponds to resolutions of \(4\), \(8\), \(16\), and \(32\) cells per droplet diameter. The study focuses on the prediction of the \(d^2\) law vaporization time constant, because this parameter is one of the few values that is easily measured in both experimental and computational studies. Following Sirignano \cite{Sirignano2010} we write
\begin{align}
  \frac{d^2}{d_0^2}=1-\frac{t}{\tau},\label{eq:d2_t}
\end{align}
where \(d_0\) is the initial droplet diameter, and \(\tau\) is the vaporization time constant. From \cref{eq:d2_t} it can be seen that \(\tau\) is the time necessary to fully vaporize a droplet. This value can be computed as
\begin{align}
  \tau = \frac{\rho_Ld_0^2}{8\rho_GD\ln(1+B)}.
\end{align}
Using the parameters of this study, the time constant is computed to be \(\tau =0.029126\text{ s}\) \cite{Sirignano2010}. Simulations are run until time \(t=1.5\times 10^{-3}\text{ s}\) when the ratio \(d^2/d_0^2\) is computed, and \cref{eq:d2_t} is used to extract a time constant. This choice of simulation end time corresponds to \(t/\tau\approx 0.01\), which is before boundary effects are important.

\begin{figure}
  \centering
  \begin{subfigure}{0.49\textwidth}
    \centering
    \includetikz{\linewidth}{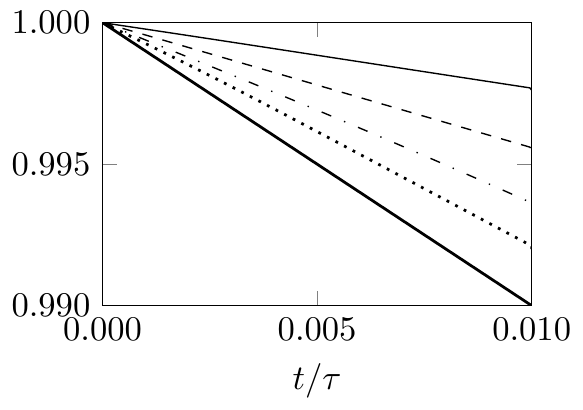}
    \caption{\(d^2/d^2_0\) in time. \(N_d=4\) (\reftikz{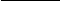}{tikz:d2slope:032}), \(N_d=8\) (\reftikz{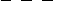}{tikz:d2slope:064}), \(N_d=16\) (\reftikz{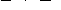}{tikz:d2slope:128}), \(N_d=32\) (\reftikz{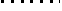}{tikz:d2slope:256}), Theory (\reftikz{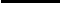}{tikz:d2slope:theory}).}\label{fig:D2_trends}
  \end{subfigure}
  \begin{subfigure}{0.49\textwidth}
    \centering
    \begin{tabular}{r c c}
      \toprule
      \(N_d\) & MB & REA2 \\
      \midrule
      \(4\)  & \(0.127952\) & \(0.125852\) \\ 
      \(8\)  & \(0.066022\) & \(0.065927\) \\ 
      \(16\) & \(0.047338\) & \(0.045630\) \\ 
      \(32\) & \(0.039449\) & \(0.036889\) \\
      \bottomrule   
    \end{tabular}
      \caption{Time constants from \(d^2\) law convergence study. The theoretical value is \(\tau=0.029126\) s.}\label{tab:D2_data}
  \end{subfigure}
\end{figure}

\Cref{fig:D2_trends} shows the temporal variation of the effective droplet square diameter, computed as
\begin{align}
  \frac{d^2}{d^2_0}={\left(\frac{V}{V_0}\right)}^{2/3},
\end{align}
where \(V_0\) and \(V\) are the initial and current droplet volumes, respectively. It can be seen that a linear trend in \(d^2\) is recovered for all cases, however, the slope of the line depends on the mesh resolution. As the grid is refined, \(\tau\) appears to approach its expected value as shown in \cref{tab:D2_data}, however, the rate appears to be less than first order accurate. This is not entirely surprising. As pointed out in \cref{sec:2D}, the evaporation rate is related to the derivative of the scalar fields, and therefore is expected to be less accurate than the scalar field which is only first order.

\begin{table}
  \centering
  \begin{tabular}{l c l l}
    \toprule
    Properties & Units & Gas & Acetone \\
    \midrule
     \(\rho\) & \(\text{kg}/\text{m}^3\) & \(1\) & \(700\) \\
     \(\mu\) & \(\text{kg}/\left(\text{m}\cdot\text{s}\right)\) & \(1\times 10^{-5}\) & \(3.26\times 10^{-4}\) \\
     \(C_p\) & \(\text{J}/\left(\text{kg}\cdot\text{K}\right)\) & \(1000\) & \(2000\) \\
     \(k\) & \(\text{W}/\left(\text{m}\cdot{K}\right)\) &  \(5.2\times 10^{-2}\) & \(1.61\times 10^{-1}\) \\
     \(D\) & \(\text{m}^2/\text{s}\) &  \textemdash & \(5.2\times 10^{-5}\) \\
     \(M\) & \(\text{kg}/\text{mol}\) & \(0.029\) & \(0.058\) \\
     \(\sigma\) & \(\text{N}/\text{m}\) & \textemdash & \(0.0237\)\\
     \(L_V\) & \(\text{J}/\text{kg}\) & \textemdash &\(5.18\times 10^{5}\) \\
     \(T_{boil}\) & \(\text{K}\) & \textemdash &\(329\) \\
    \bottomrule
  \end{tabular}
  \caption{Fluid properties for \(d^2\) law problem.}\label{tab:D2}
\end{table}

%% \begin{table}
%%   \centering
%%   \begin{tabular}{r c c}
%%     \toprule
%%     \(N_d\) & MB & REA\\
%%     \midrule
%%     \(4\)  & \(0.127952\) & \(0.125852\) \\ 
%%     \(8\)  & \(0.066022\) & \(0.065927\) \\ 
%%     \(16\) & \(0.047338\) & \(0.045630\) \\ 
%%     \(32\) & \(0.039449\) & \(0.036889\) \\
%%     \bottomrule   
%%   \end{tabular}
%%   \caption{Time constants from \(d^2\) law convergence study. Rate of convergence computed relative to preceding simulation. The theoretical value is \(\tau=0.029126\) s.}\label{tab:D2_data}
%% \end{table}
\section{Vaporizing Droplet in Uniform Flow}\label{sec:cross}
A final test is performed of a vaporizing droplet in uniform flow. There is no known analytical solution to this problem. A droplet is placed at the center of a gaseous domain with inflow velocity \(U=40\) m/s. Unlike the simulation of the previous section, the assumption \(\bm u_L=0\) is not imposed, so the droplet is able to move and deform freely in space. The same fluid properties are used as the \(d^2\) law case, \cref{tab:D2}. The solution is initialized with a \(d^2\) law initialization. The parameters chosen lead to a flow Reynolds number \(Re=\rho_GUd/\mu_G=400\) and gas Schmidt and Prandtl numbers as \(Sc=\mu_G/\left(\rho_GD\right)=0.192\text{ and }Pr=\mu_G/\left(\rho_G\lambda_G\right)=0.192\), respectively. The surface tension is not assumed to be zero, and is given as \(\sigma = 0.0237\) N/m. This leads to a Weber number \(We=\frac{\rho_G U^2d}{\sigma}\approx 7\).

The simulation used \(128\) grid cells in each direction, which was chosen to resolve the gas film thickness, \(\delta\). Following the paper by Sirignano and Abramzon \cite{Abramzon1989}, the following correlation for gas film thickness can be derived,
\begin{align}
  \delta = \frac{d}{0.552{Re}^{1/2}{\max\left(Sc,Pr\right)}^{1/3}}. \label{eq:thickness}
\end{align}
For the parameters in this study, \cref{eq:thickness} yields \(\delta\approx 0.157 d\) or \(\delta/\Delta\approx 2.51\). \Cref{eq:thickness} is noted to underpredict the film thickness \cite{Abramzon1989}, so the use of \cref{eq:thickness} can be considered a upper bound on the required resolution.

\begin{figure}
  \centering
  \begin{subfigure}{0.49\textwidth}
    \includegraphics[height=\linewidth,angle=270]{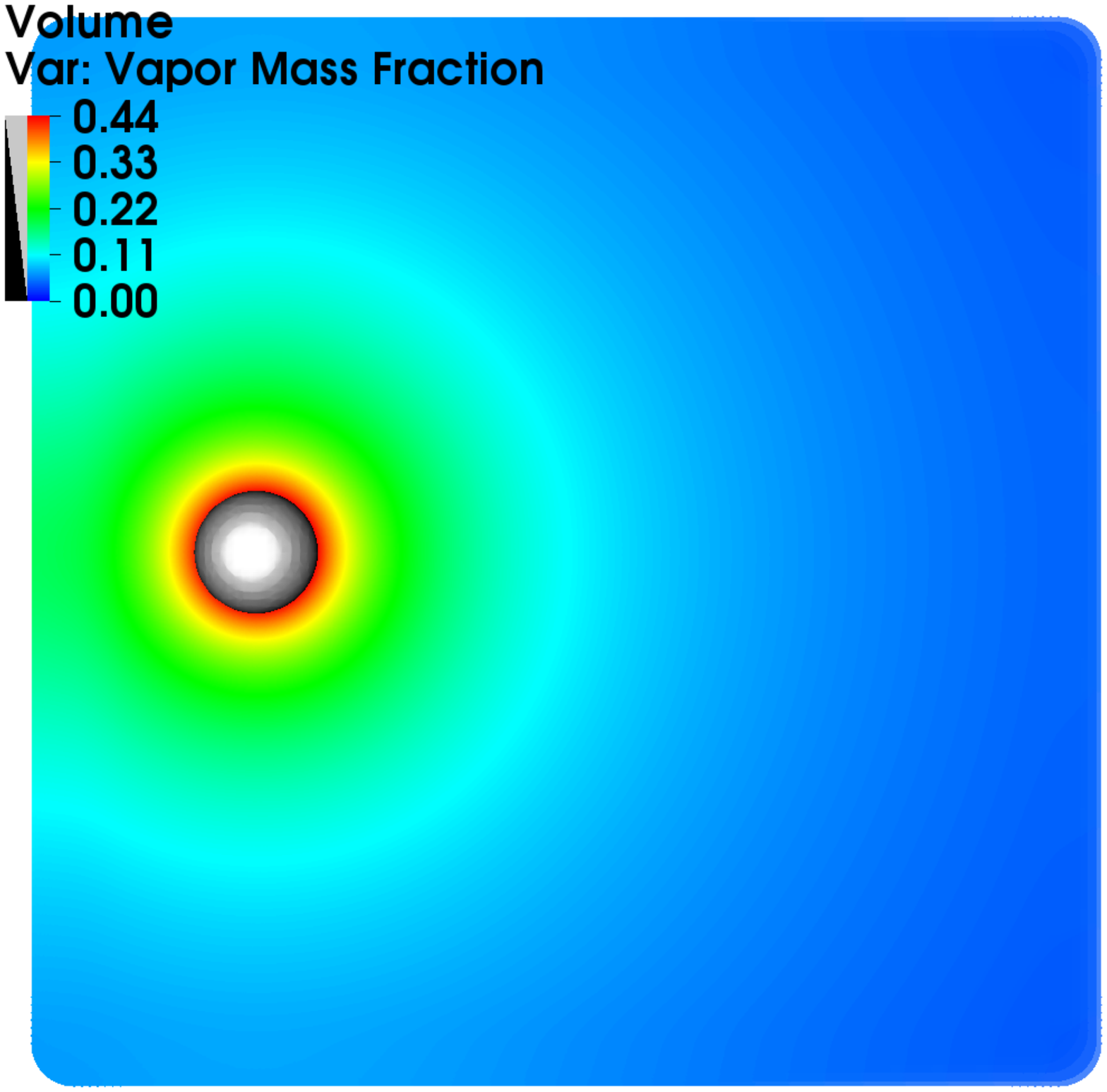}
    \caption{Initial condition (\(t/\tau=0\)).}\label{fig:cross_1}
  \end{subfigure}
  \begin{subfigure}{0.49\textwidth}
    \includegraphics[height=\linewidth,angle=270]{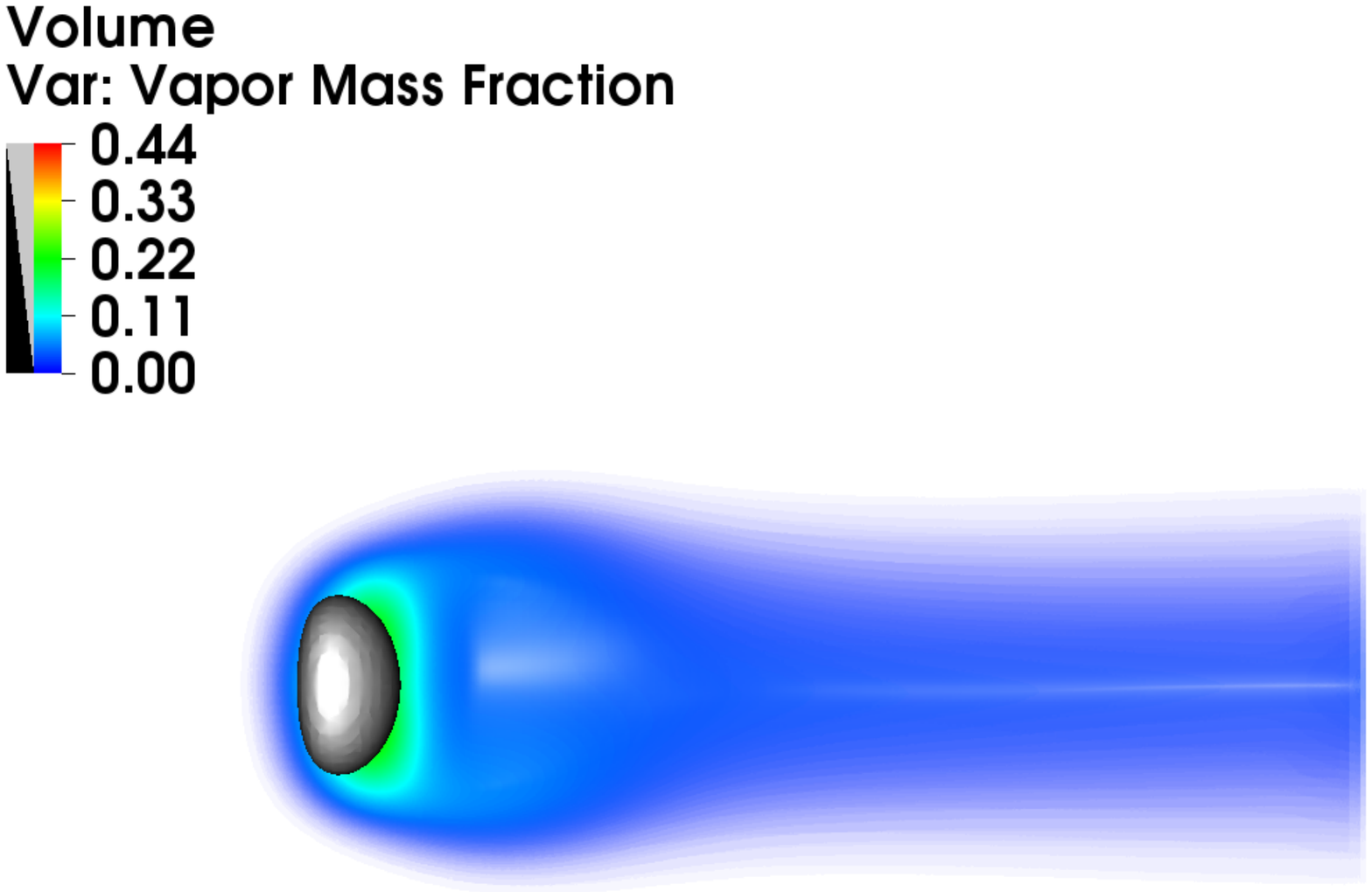}
    \caption{\(t/\tau=7.21\text{e-}5\).}\label{fig:cross_2}
  \end{subfigure}
  \begin{subfigure}{0.49\textwidth}
    \includegraphics[height=\linewidth,angle=270]{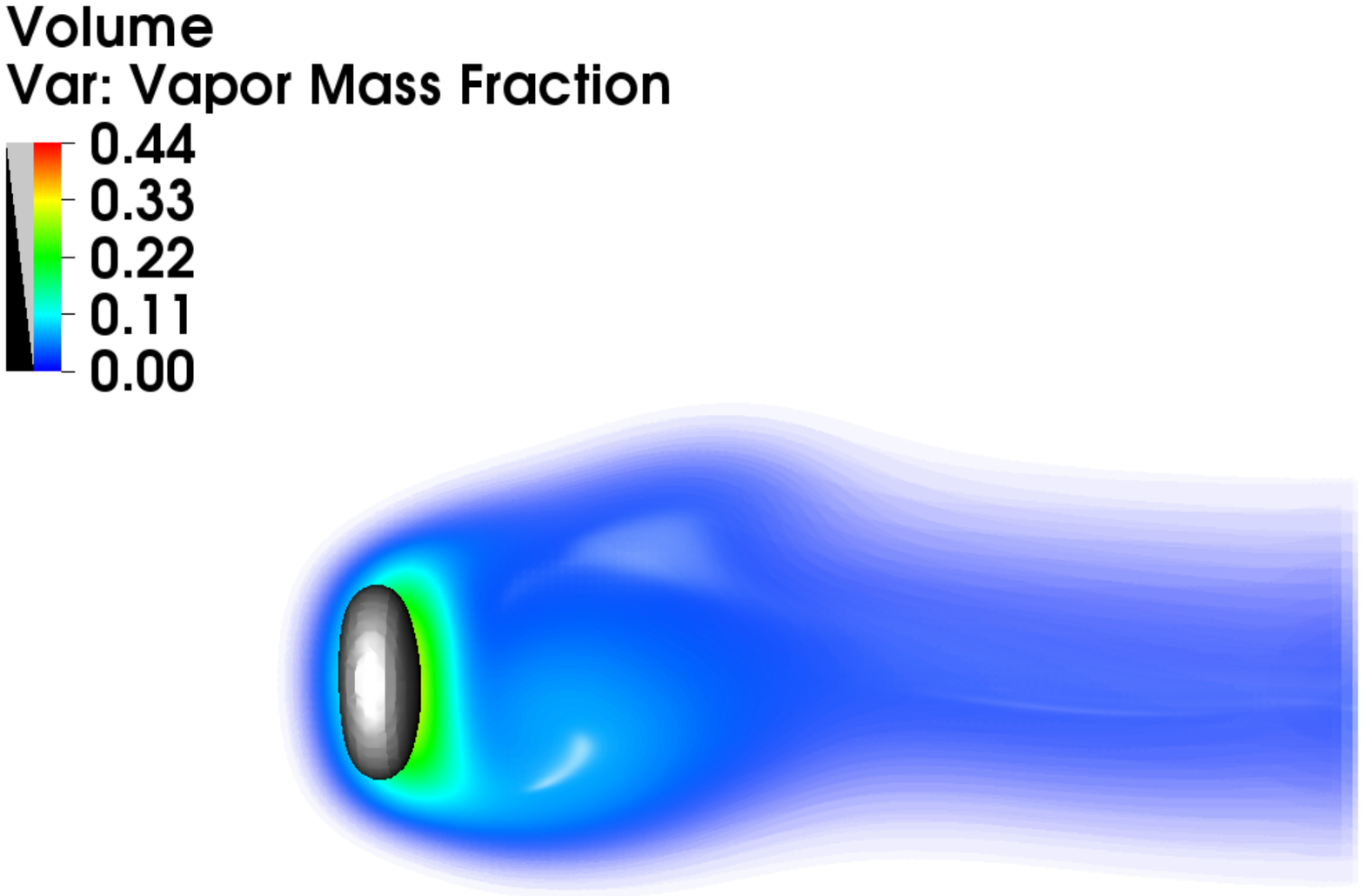}
    \caption{\(t/\tau=1.10\text{e-}4\).}\label{fig:cross_3}
  \end{subfigure}
  \begin{subfigure}{0.49\textwidth}
    \includegraphics[height=\linewidth,angle=270]{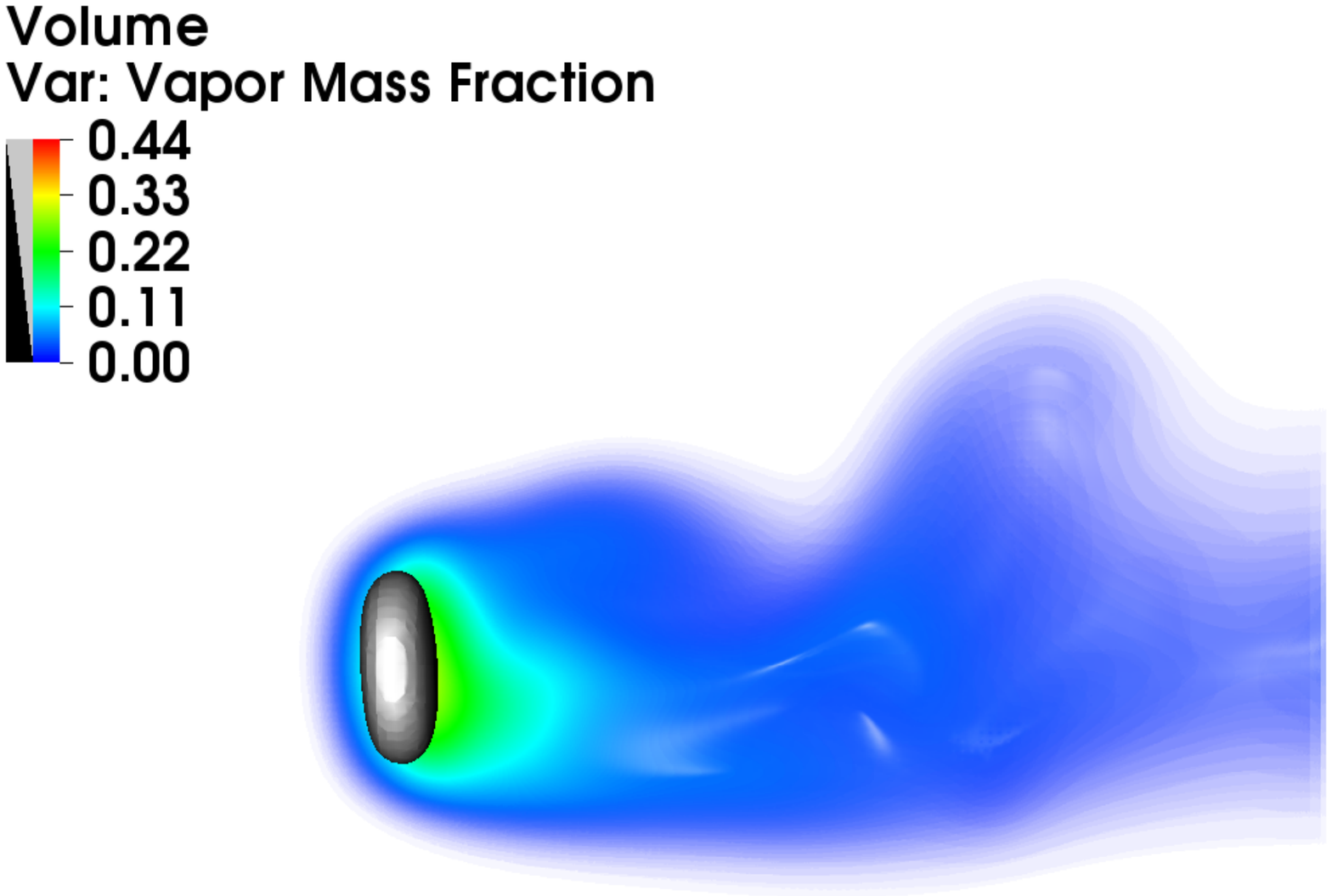}
    \caption{\(t/\tau=1.34\text{e-}4\).}\label{fig:cross_4}
  \end{subfigure}
  \caption{Volume render of vapor mass fraction at representative times. The rendering is truncated at the midplane in order to visualize the interface (shown in gray).}\label{fig:cross_series}
\end{figure}

\begin{figure}
  \centering
  \includetikz{0.49\textwidth}{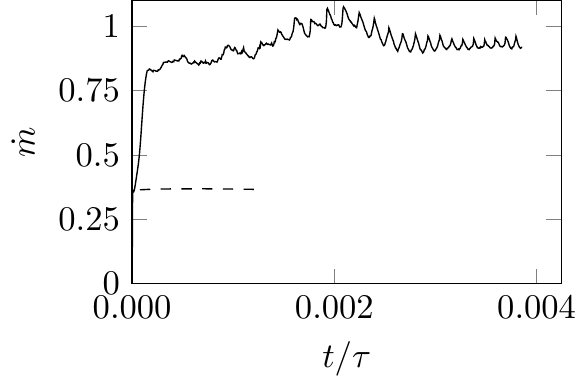}
  \caption{\(\dot m\) in time. \(Re=400\) (\reftikz{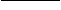}{tikz:crossmdot:flow}), \(Re=0\) (\reftikz{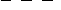}{tikz:crossmdot:static}).}\label{fig:cross_mdot}
\end{figure}

A series of visualizations is shown in \cref{fig:cross_series}. \Cref{fig:cross_1} demonstrates the initial condition. As the flow progresses, aerodynamic forces cause the droplet to flatten and move slightly in the domain (\cref{fig:cross_2}). \Cref{fig:cross_3} shows the wake becoming asymmetrical and the formation of a large vortex under the droplet. After some time, the droplet wake becomes chaotic due to the interaction of several shed vortices, \cref{fig:cross_4}. The temporal variation in vaporization rate is shown in \cref{fig:cross_mdot}. The vaporization rate is greatly enhanced by the flow when compared to the \(Re=0\) limit (the \(d^2\) law). The vaporization rate appears to increase in time until the attached vortices begin to shed, indicating strong coupling between vortex formation and vaporization.

\section{Conclusion}
A numerical solver capable of simulating droplet vaporization in a complex flow field is demonstrated. A discretely conservative volume of fluid scheme is coupled with appropriate sources based on local thermodynamic equilibrium of the vapor mass fraction and temperature fields. The evaporation sources are coupled to the scalar transport equation using an unconditionally stable, monotone scheme. The convergence and stability properties of the solver are explored, and it shown that first order convergence is seen for gas temperature and vapor mass fraction, consistent with the schemes used. Also consistent with the scheme accuracy is the fact that the liquid velocity seems to converge poorly. Future work will require the use of higher order schemes for the scalar transport, to avoid this issue. Finally, it is shown that when used with a fully implicit time integrator the solver is capable of running stably at high time steps (for example \(C_\lambda=200\)) with non-oscillatory solutions. This is an improvement over the common Crank-Nicolson time integrator, and the improvement comes without any sacrifice of accuracy.

As true test of the robustness of the numerical solver, three dimensional simulations at high Reynolds number are performed. A vaporizing droplet in uniform flow at \(Re=400\) is demonstrated. Because of the high speed flow, there is significant deformation of the interface, so no analytical solution is known for this problem. It is noted that vaporization is enhanced by the flow, which is consistent with the literature on non-deforming vaporizing droplets \cite{Abramzon1989}. %% Secondly, a simulation is shown of a \(Re=10000\) atomizing liquid jet undergoing vaporization, demonstrating the robustness of the solver to both flow turbulence and complex changes in interface topology.
\section{Acknowledgments}
This material is based upon work supported by the National Science Foundation Graduate Research Fellowship under Grant No.~DGE-1650441 and by the Alfred P.~Sloan Foundation. This work used the Extreme Science and Engineering Discovery Environment (XSEDE), which is supported by National Science Foundation grant number ACI-1548562. The XSEDE resource used was Stampede2 at the Texas Advanced Computing Center (TACC) at The University of Texas at Austin. The authors also acknowledge Advanced Research Computing at Virginia Tech for providing computational resources and technical support that have contributed to the results reported within this paper. 
\section{Appendix}
\subsection{Monotone Schemes}\label{sec:monotoneProof}
Harten \cite{Harten1983} in the development of total variation diminishing (TVD) schemes defines three types of schemes. Monotonicity preserving schemes, TVD schemes, and monotone schemes. Monotonicity preserving schemes are those that do not introduce new extrema to the solution, nor do they change the value of the current extrema. TVD schemes are defined in terms of a mathematical property called the total variation, and they are shown to be a class of monotonicity preserving schemes. The final category, monotone schemes, are a robust class of TVD schemes, and therefore also preserve monotonicity. Following Harten, define an update operator such that
\begin{align}
  \bm\phi^{n+1}=\bm{\mathcal L}\bm\phi^{n},\label{eq:TVD}
\end{align}
where the vector \(\bm\phi=[\phi_1,\phi_2,\ldots]^\top\). A monotone scheme is one such that the coefficients of \(\bm{\mathcal L}\) are all positive. Comparing \cref{eq:monotone_discrete} to \cref{eq:TVD}, the chosen time-space integration scheme is monotone when the matrix,
\begin{align}
  \bm{\mathcal A}=\left\{
  \begin{array}{lc}
    A_{i,j}=1+2C & j=i \\
    A_{i,j}=-C & j=i-1 \text{ or } j=i+1\\
    A_{i,j}=0 & \text{otherwise,}
  \end{array}
  \right.
\end{align}
is invertible, and the coefficients of \({\bm{\mathcal A}}^{-1}\) are all positive. This property follows directly from the fact that \(\bm{\mathcal A}\) is a nonsignular M matrix\cite{Plemmons1977}. That is, \(\bm{\mathcal A}\) is an M matrix, and it is strictly diagonally dominant with positive diagonal.

\subsection{Farfield Boundary}\label{sec:far}
\begin{figure}
  \centering
  \includetikz{0.618\textwidth}{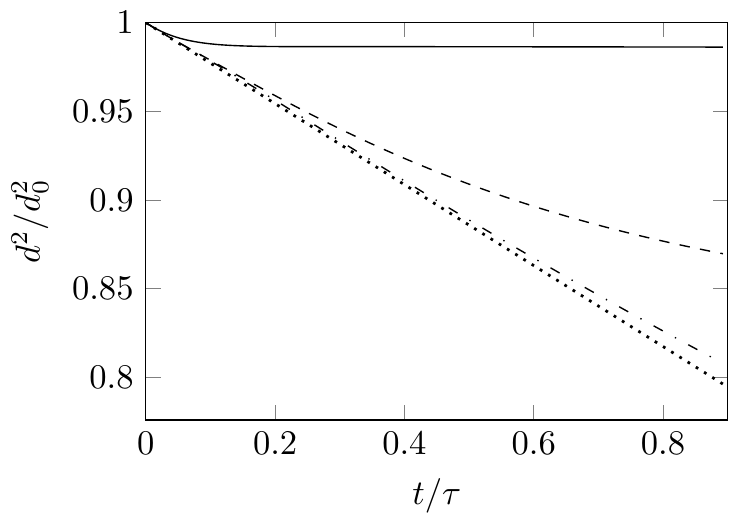}
  \caption{\(d^2\) scaling as a function of domain size. Domain size varies as
four diameter (\reftikz{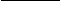}{tikz:bound:04}), eight diameters (\reftikz{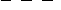}{tikz:bound:08}), and sixteen diameters (\reftikz{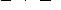}{tikz:bound:16}). An extrapolated linear trend is shown as (\reftikz{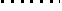}{tikz:bound:trend}).}\label{fig:boundary}
\end{figure}

The \(d^2\) law is the vaporization of an isolated droplet in a
quiescent field, which calls for an infinite domain. In this work, the domain is
approximated using a triply periodic cube, with a fixed droplet in the
center. The domain is made to be much larger than the droplet so that the
effects of the boundary will be minimal. Still, as the finite sized domain
fills with vapor, the effects of the boundary condition will be felt. A study
was performed to look at the impact of domain size (measured in droplet
diameters) on the evaporation rate. The droplet diameter, fluid properties, and
initial conditions were made to be consistent with those in \cref{sec:3D}. For
this study, the grid resolution is fixed so that the number of grid cells per
droplet diameter is always \(8\). The domain sizes were chosen to be integral
multiples of the droplet diameter, viz., \(4\), \(8\), and \(16\) diameters. The
results of this study can be seen in \cref{fig:boundary}. At the beginning of
the simulation, all domains give the same trend in \(d^2\). As the simulation
progresses, boundary effects become important. This manifests itself in a gradual plateau of the droplet square diameter as a function of time. Simulations with a greater ratio of domain size to droplet diameter are able to continue longer without this plateau. As a reference, a straight line is shown by extrapolating the slope from the initial condition to the first time step. This reference was chosen instead of the analytical result, \(1/\tau\), since the effective vaporization rate, and therefore the slope, changes under mesh refinement. For the given parameters, a domain eight times the droplet diameter leaves an error of only about \(1.3\%\) in the slope at \(t=0.01\tau\), which is the standard used in \cref{sec:3D}.
\section{References}
\bibliographystyle{ieeetr}
\bibliography{report}
\end{document}